\documentclass[%
reprint,
nofootinbib,
 amsmath,amssymb,
aps,
]{revtex4-1}

\usepackage{tikz}
\usetikzlibrary{matrix,arrows,decorations.pathmorphing,decorations.pathreplacing}
\usepackage{euscript,array,mathrsfs}

\def\BP{P}

\def\BS{S}
\def\BA{A}

\def\BH{{\mathscr H}}

\newcommand{\Tr}{\operatorname{Tr}}
\newcommand{\IM}{\operatorname{Im}}
\newcommand{\RE}{\operatorname{Re}}

\newcommand{\SUP}{\operatorname{sup}}

\def\bra#1{\langle #1|}
\def\ket#1{| #1\rangle}

\def\PP{\mathscr{P}}

\newcommand{\EQ}[1]{\begin{equation}\begin{split} #1
\end{split}\end{equation}}

\newcommand{\BOX}[1]{
\begin{center}\fbox{\parbox{8cm}{#1}}\end{center}}

\begin{document}

\preprint{APS/123-QED}

\title{The Emergent Copenhagen Interpretation of Quantum Mechanics}

\author{Timothy J. Hollowood}
\email{t.hollowood@swansea.ac.uk}
\affiliation{
Department of Physics, Swansea University,\\ Swansea, SA2 8PP, UK
}

\date{\today}

\begin{abstract}
We introduce a new and conceptually simple interpretation of quantum mechanics based on reduced density matrices of sub-systems from which the standard Copenhagen interpretation emerges as an effective description of macroscopically large systems. Wave function collapse is seen to be a useful but fundamentally unnecessary piece of prudent book keeping which is only valid for macro-systems. 
The new interpretation lies in a class of modal interpretations in that it applies to quantum systems that interact with a much larger environment. However, we show that it 
does not suffer from the problems that have plagued similar modal interpretations like 
macroscopic superpositions and rapid flipping between macroscopically distinct states. We describe how the interpretation fits neatly together with fully quantum formulations of statistical mechanics and that 
a measurement process can be viewed as a process of ergodicity breaking analogous to a phase  transition. The key feature of the new interpretation is that joint probabilities for the ergodic subsets of states of disjoint macro-systems only arise as emergent quantities. 
Finally we give an account of the EPR-Bohm thought experiment and show that the interpretation implies the violation of the Bell inequality characteristic of quantum mechanics but in a way that is rather novel. The final conclusion is that the  Copenhagen interpretation gives a completely satisfactory phenomenology of macro-systems interacting with micro-systems.
\end{abstract}

\maketitle


\section{Introduction}\label{s1}

The central mystery of quantum mechanics is present even in the simplest measurement on a qubit. Solving the Schr\"odinger equation for a suitable Hamiltonian gives an evolution of the form
\EQ{
&\big(c_+\ket{z^+}+c_-\ket{z^-}\big)\otimes\ket{A_0}\\
&\qquad~~~~~~~~~\longrightarrow c_+\ket{z^+}\otimes\ket{A_+}+c_-\ket{z^-}\otimes\ket{A_-}\ ,
\label{fzz}
}
where the two states $\ket{A_\pm}$ are macroscopically distinct states of the measuring device. But how can this be consistent with the fact that when any experiment of this type is performed a definite outcome occurs either $\ket{A_+}$ or $\ket{A_-}$? The Copenhagen interpretation\footnote{The Copenhagen interpretation is not really a completely settled set of ideas. We are using the term to stand for the way that most working physicists successfully use quantum mechanics in practice without having to even think about foundational issues.} solves the problem by ``collapsing the wave function," i.e.~choosing only one of the distinct terms on the right-hand side with probabilities $|c_\pm|^2$, respectively, on the grounds that the measuring device is macroscopic. The universal success of applying the rule disguises the fact that it is really only a rule of thumb: when is a system sufficiently macroscopic that it qualifies for collapse? This arbitrary separation of systems into microscopic and macroscopic is the famous Heisenberg cut.

We can measure how macroscopically distinct the two states of the measuring device $\ket{A_\pm}$ are by estimating their inner product. Let us suppose the measuring device has a macroscopically large number of microscopic degrees-of-freedom $N$. States are macroscopically distinct if all the microscopic degrees-of-freedom 
are separated by a macroscopic scale $L$. If $\ell$ is a characteristic microscopic length scale in the system and assuming, say, Gaussian wave functions for the microscopical degrees-of-freedom spread over the scale $\ell$, the matrix elements between macroscopically distinct states is roughly
\EQ{
\Delta\equiv\big|\bra{A_+}A_-\rangle\big|\thicksim \exp\big[-NL^2/\ell^2\big]\ .
\label{est2}
}
Just to get a feel for the numbers, suppose $N\sim10^{20}$, $\ell\sim10^{-10}\,\text{m}$ (atomic size) and $L\sim10^{-4}\,\text{m}$, giving
\EQ{
\boxed{\Delta\thicksim e^{-10^{32}}}
\nonumber
}
In the following we will use $\Delta$ to denote a generic scale characterising the inner products of macroscopically distinct states. The estimate above is intended as a guide and the
fact that this generic scale is so small will play an important role in this work.

There is an interesting analogue of the measurement problem in classical statistical mechanics.
Consider an Ising ferromagnet above its Curie temperature. In a typical configuration,
 the spins point randomly up or down and there is no net magnetization. In the standard interpretation of classical statistical mechanics, the ensemble average captures a time average of the dynamics of the underlying microscopic state of the system. This is a statement of ergodicity: over time, interactions ensure that 
the microscopic state explores all the available states with a probability given by the usual Boltzmann factor. Now, if the temperature is reduced below the Curie temperature, there is a phase transition and the magnet finds itself in an ordered state where the spins tend to line up in the same direction, either up or down, and the system develops a net magnetization.
At a microscopic level, ergodicity is broken and for a given initial micro-state the system effectively only explores half of the overall ensemble over time and it is each of these sub-ensembles that has a net magnetization.

This picture of ergodicity breaking provides a possible template for solving the quantum measurement problem. The key feature that allows for its breaking in classical statistical mechanics is that the 
state of the system has two dual aspects: the ensemble and the micro-state. Of course, the fact that we deal with ensembles is because we, as macroscopic observers, are ignorant as to the exact micro-state at any particular time.
When the system is in equilibrium, microscopic  interactions in the system ensure that the micro-state explores the set of available states ergodically and a suitably coarse-grained time average is equal to the ensemble average.
However, when the magnet undergoes the phase transition to the low temperature phase, the system becomes highly sensitive to which of the sub-ensembles the micro-state lies in. After the transition, ergodicity is broken and the micro-state only explores a sub-ensemble over time and so the time average is captured by the sub-ensemble. What is crucial for this mechanism is that underlying the ensemble is the 
existence of a real micro-state, even though this is hidden from the macroscopic point-of-view.

 In order to carry this over into the quantum mechanical measurement problem, we need the analogues of the 
 ensemble and the micro-state or what is called an ``epistemic state" and an ``ontic state".\footnote{The latter were called
``internal states" in \cite{Hollowood:2013cbr} after \cite{Bene:1997kk,Sud1}.}
\EQ{
\text{\bf quantum mech.} ~&\qquad ~\text{\bf statistical mech.}\\
\text{epistemic state}~~~ &\leftrightsquigarrow ~~~~~~~\text{ensemble}\\
\text{ontic state} ~~~~~~&\leftrightsquigarrow \,~~~~~\text{micro-state}\nonumber
}
If we had such a dual aspect, or different {\it modalities\/}, of the quantum state then we
would have the possibility of applying an argument that involves the breaking of ergodicity to the measurement problem. 

We do not expect that the macroscopic measuring device $A$ will be in a pure quantum state, rather, it will be described by a reduced density matrix $\hat\rho_A$ as a consequence of its interaction with its environment. This density matrix is the epistemic state, the analogue of the ensemble in classical statistical mechanics. However, quantum mechanics in the standard formulation offers no analogue of an ontic state. In fact
any attempt to provide an additional specification of the quantum state is tantamount to the introduction of hidden variables and, as shown by Bell \cite{Bell:1964oeprp}, generally leads to
predictions that are not consistent with standard quantum mechanics and experiments.
So the conclusion is that classical statistic mechanics cannot provide a perfect analogue of what happens in a quantum measurement unless there is some way to introduce addition information into the quantum state, the ontic information, but in a way that would be consistent with the usual predictions of quantum mechanics including the violation of Bell's theorem.
A clue to how this could be done, is provided by a class of {\it modal interpretations\/} of quantum mechanics. In some of these interpretations, a system, being a sub-system of a larger system, is described by a reduced density matrix, the epistemic state. But as well as having a density matrix, the system also carries an ontic state in the form of one of the eigenvector of the reduced density matrix. These are the two modalities of the quantum state.
\begin{figure}[ht]
\begin{center}
\begin{tikzpicture}[xscale=0.8,yscale=0.8]
\draw (0,0) ellipse  (3cm and 2cm);
\draw (-1,-0.5) ellipse  (1cm and 0.7cm);
\node at (-0.5,-0.3) (a1) {$A$};
\node at (0.9,0.7) (a1) {$E$};
\node at (-3.9,1.7) (a1) {$S:~~\ket{\Psi}$};
\node at (-1,-2.5) (a2) {$\hat\rho_A=\Tr_E\ket{\Psi}\bra{\Psi}=\sum_ip_i\ket{\psi_i}\bra{\psi_i}$};
\draw[->] (-3.4,-2.2) -- (-1.2,-0.7);
\draw[->] (-3.4,1.4) -- (-2.8,0.8);
\end{tikzpicture}
\end{center}
\caption{\small A sub-system $A$ of a large quantum system $S=A+E$. The quantum state of $A$ is described by the reduced density matrix $\hat\rho_A$. In the new interpretation the ontic state of $A$ at a given time corresponds to one of the eigenvectors of $\hat\rho_A$.}
\label{f1}
\end{figure}
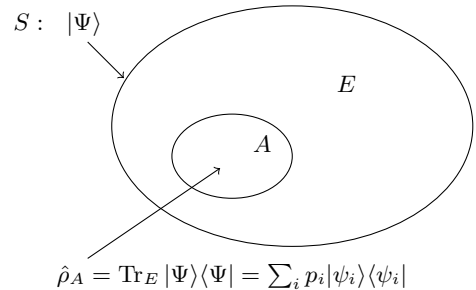
Once we allow the system to have its own ontic state then this opens up the possibility to solve the measurement problem by using an ergodicity argument along the lines of classical statistical mechanics.\footnote{The fact that ergodicity breaking should play some role in the measurement 
problem crystalized in discussions with Jacob Barandes.} The question that we address in this paper is whether a complete and consistent interpretation of quantum mechanics can be constructed along these lines. Our claim---building on \cite{Hollowood:2013cbr}---is that there exists a conceptually simple interpretation that can  
fill the ugly gap in the Copenhagen interpretation between unitary time evolution of states and the collapse of the wave function by having these features apply to the two different modes of the quantum state.
At the same time it appear to evade the consequences of Bell's theorem and its generalizations to be fully consistent with the usual predictions of quantum mechanics.
We call this new interpretation the {\it emergent Copenhagen interpretation\/} for reasons that will themselves emerge. It lies firmly in the class of modal interpretations \cite{Krips:1969tpqm,Krips:1975siqt,Krips:1987mqt,vanFraassen:1972faps,Cartwright:1974vfmmqm,Krips:1987mqt,vanFraassen:1991qmmv,Bub:1992qmwpp,VermaasDieks:1995miqmgdo,BacciagaluppiDickson:1999dmi}\footnote{The book by Vermaas \cite{Vermaas:1999puqm} gives an eloquent account of the class of modal interpretations, both their successes and problems, as well as having a full list of references.} but is rather different in some fundamental ways that ensure it does not suffer from the problems that plague other modal interpretations.

The new interpretation, like other modal interpretations, works hand-in-hand with the phenomenon of decoherence in quantum systems \cite{Bohm:1951qt,JoosZeh:1985ecptiwe,Joos:2003dacwqt,Zurek:2003dtqc,Schlosshauer:2005dmpiqm,SchlosshauerCamilleri:2008qct,BreuerPetruccione:2002toqs} in order to describe a satisfactory phenomenology of macroscopic systems and is closely allied to the modern understanding of statistical mechanics as it arises from the quantum mechanics of systems interacting with large environments or baths \cite{PopescuShortWinter:2005fsmeisa,PopescuShortWinter:2006efsm,BL,LL,GoldsteinLebowitzTumulkaZanghi:2006ct,GMM,LPSW,Sh}. Classical behaviour, the Born rule and wave function collapse are not put in by hand but are seen to emerge as an approximate phenomenology of systems with a macroscopic number of degrees-of-freedom. Heisenberg's cut is replaced by a continuous spectrum of ``classicality".
Finally, the new interpretation  leads to a solution of the measurement problem via the mechanism of ergodicity breaking.

The paper is organised as follows. In section \ref{s1.1} we introduce the new interpretation by defining ontic states in section \ref{s1.11}. In section \ref{s1.12}, we discuss disjoint sub-systems and how their ontic states can relate to the ontic states of their union. In section \ref{s1.2}, we discuss sub-systems in general and the extent to which quantum descriptions of a system are to be thought of as effective theories with an in-build ultra-violet cut off. Section \ref{s1.3} then considers the dynamics of ontic states. This takes the form of a stochastic process which must satisfy a number of conditions. Most importantly, as we discuss in section \ref{a1}, it must be coarse-grained at the scale of the ultra-violet cut off in the temporal domain to avoid problems of other modal interpretations. Section \ref{s2} discusses how a recognisable classical ontology can emerge for macro-systems. This involves a discussion of how a classical ontology involves a patching together of ontic states of a number of macro-systems embedded in a larger environment. We show that it is meaningful to define joint probabilities for disjoint systems but only in an emergent sense. Section \ref{s2.9} explains how the new interpretation is related to modern formulations of statistical mechanics built on quantum mechanics. Section \ref{s3} is devoted to a discussion of measurement. We first show in section \ref{s2.1} how a simple model without an environment can account for some features of measurement but also has a number of problems. In section \ref{s2.2}, these problems are all resolved in a more realistic model that includes the environment as well as a measuring device that is not 100\% efficient. In section \ref{s4.1} we discuss the classical experiment of Bohm based on the original thought experiment of Einstein, Podolsky and Rosen and show the new interpretation gives a description that reproduces that of Copenhagen quantum mechanics but without the need to collapse the wave function. Finally, in section \ref{s5}, we draw some conclusions.

\section{The Emergent Copenhagen Interpretation }\label{s1.1}

\subsection{Ontic States}\label{s1.11}

A key feature of the new interpretation, as in some other modal interpretations, is the focus on sub-systems of larger systems 
$\BA\subset\BS$: see figure \ref{f1}.
For simplicity, we assume that $S$ is large enough so that its quantum dynamics is to an excellent approximation unitary. In most cases, we can assume that the state of $S$ is pure $\ket{\Psi(t)}$; indeed, much of the 
what we say will be independent of the exact state of $S$ whether mixed or pure.\footnote{Note that the mixed states that we have are all ``improper mixtures" and we do not need the concept of a ``proper mixture".}
For each sub-system, $\BA\subset\BS$ for which the Hilbert space of $\BS$ factorizes as $\BH_\BS=\BH_\BA\otimes\BH_E$, we can define a reduced density matrix by tracing over the Hilbert space of the complement $E$:
\EQ{
\hat\rho_A(t)=\text{Tr}_E\ket{\Psi(t)}\bra{\Psi(t)}\ .
}
It is a theorem that the reduced density $\hat\rho_A(t)$ has a discrete spectrum:
\EQ{
\hat\rho_A(t)\ket{\psi_i(t)}=p_i(t)\ket{\psi_i(t)}\ ,
\label{pdx}
}
where the $\{p_i(t)\}$ are a set of real numbers with $0\leq p_i(t)\leq1$ and $\sum_ip_i(t)=1$. The reduced density matrix $\hat\rho_A$ is the {\it epistemic state\/} of $A$.

\BOX{{\bf The ontic state:} at a particular instant of time, the new interpretation asserts that $\BA$ is actually in one of the eigenstates $\ket{\psi_i(t)}$. The state that is actually realised is called the {\it ontic state\/}.
}
This property assignment is done at an instant of time $t$ and therefore a degeneracy in the $\{p_i(t)\}$ is not realistic. However, dealing with problems that arise from degeneracies, or more realistically near-degeneracies, as time evolves is key to building a successful modal interpretation. Note, also that the time-dependence of the ontic state $\ket{\psi_i(t)}$ refers to the time at which the decomposition \eqref{pdx} is made and it is important that these states do not generally solve the Schr\"odinger equation.\footnote{In fact, since $A$ is generally interacting with $E$ there is no concept of a Schr\"odinger equation applying within the sub-system $A$.}
 
Unlike other modal interpretations, we do not directly interpret $p_i(t)$ as the probability that $A$ is in the ontic state $\ket{\psi_i(t)}$, although this will emerge in certain situations.
In fact, the more fundamental probabilities in the new interpretation are conditional probabilities $p_{i|j}(t,t_0)$ that, given the system was in the ontic state $\ket{\psi_j(t_0)}$ at an earlier time $t_0<t$, the system is in the ontic state $\ket{\psi_i(t)}$ at time $t$. It is a hypothesis that these conditional probability are related to the single-time probabilities $p_i(t)$ via
\EQ{
\boxed{p_i(t)=\sum_jp_{i|j}(t,t_0)p_j(t_0)}
\label{pc6}
}
Given this, there are two ways that the $p_i(t)$ can be interpreted as single-time probabilities:
\begin{description}
\item[Initial condition] the key equation \eqref{pc6} shows that is an
unambiguous way to define the single-time probabilities $p_i(t)$ via an initial condition.
If at $t_0$ the state of $S$ is a tensor product state $\ket{\psi_0(t_0)}\otimes\ket{\phi_0(t_0)}$, then the ontic state at $t_0$ is uniquely $\ket{\psi_0(t_0)}$. In that case, $p_i(t)=p_{i|0}(t,t_0)$ is the probability that the ontic state is $\ket{\psi_i(t)}$ at time $t>t_0$. 
\item[Equilibrium] we will see, in section \ref{s2.9}, that there is another definition that is valid when $A$ is a macro-system in equilibrium with its environment $E$ so that $\hat\rho_A(t)$ is only slowly varying. In this case, for a characteristic time scale $\tau$, $p_{i|j}(t,t_0)$, with $t=t_0+\tau$, becomes independent of $j$, the initial state. In that case, $p_i(t)=p_{i|j}(t,t_0)$ (approximately time-independent) is the probability of finding the system in the ontic state $\ket{\psi_i(t)}$, independent of the initial state $\ket{\psi_j(t_0)}$.
\end{description}

One might imagine that the assignment of a particular ontic state is tantamount to specifying a kind of hidden variable. As we will see this is potentially misleading because the behaviour of ontic states is not at all like standard hidden variables. In particular, the ontic states of $A$ are not global property assignments and the recognition that they are only intrinsic properties from the perspective of $A$ in relation to the rest of the total system
is known in the literature on modal interpretations as {\it relationalism\/} or {\it perspectivalism\/} \cite{BH,Rov,Dk1,BD1}. We want to emphasize that this is not really a philosophical stance but is 
simply acknowledging what it means to perform a trace that involves summing over states in a disjoint sub-system to $A$.

There is an important additional detail to mention here. 
Since the total system $A+E$ is in a pure state $\ket{\Psi}$, then assuming $d_A\leq d_E$, where $d_A=\text{dim}\,\BH_A$, etc., because 
the environment $E$ is typically much bigger than the sub-system of interest $A$, 
each ontic states of $A$, say $\ket{\psi_i}$, is precisely correlated with a particular ontic state of $E$, which we call the {\it mirror ontic state\/} and denote $\ket{\tilde\psi_i}$. This follows because 
$\hat\rho_A$ and $\hat\rho_E$ have the same non-vanishing spectrum and one way to exhibit the correlation is via the Schmidt decomposition of $\ket{\Psi}$:
\EQ{
\ket{\Psi}=\sum_i\sqrt{p_i}\ket{\psi_i}\otimes\ket{\tilde\psi_i}\ .
\label{m34}
}
The states $\ket{\psi_i}\otimes\ket{\tilde\psi_i}$ are a set of $d_A$ orthogonal vectors in the $d_Ad_E$ dimensional Hilbert space of $A+E$. So the property assignment of $\ket{\psi_i}$ is always precisely correlated with the mirror assignment of $\ket{\tilde\psi_i}$ to its complement. We will see in section \ref{s1.3} that the mirror ontic states play an important role in the dynamics of ontic states.

Finally, we should point out that ontic states are irreducible in the sense that there is no further notion of probability on top of their inherent probability. In this regard, we do not assume a priori the Born rule which we will have to ultimately derive from the behaviour of ontic states in realistic situations.

\subsection{Disjoint Sub-systems}\label{s1.12}

It is important that once we trace down to the Hilbert space $\BH_A$ factor, we potentially forgo any knowledge of the ontic states of other disjoint sub-systems. Generally working with the sub-space $A$ is only good for asking inclusive questions regarding the dynamics of ontic states in a disjoint sub-system. In particular, as mentioned in the last section, this means that ontic assignments for $A$ cannot generally be taken as global property assignments. 

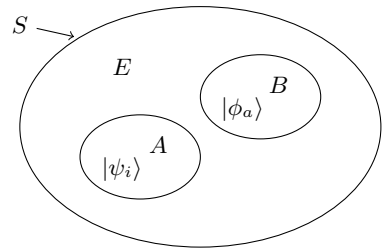
\begin{figure}[ht]
\begin{center}
\begin{tikzpicture}[xscale=0.8,yscale=0.8]
\draw (0,0) ellipse  (3cm and 2cm);
\draw (-1,-0.5) ellipse  (1cm and 0.7cm);
\draw (1,0.5) ellipse  (1cm and 0.7cm);
\node at (-0.7,-0.3) (a1) {$A$};
\node at (0.7,0.3) (a1) {$\ket{\phi_a}$};
\node at (-1.3,1) (a1) {$E$};
\node at (-3,1.7) (a2) {$S$};
\draw[->] (a2) -- (-2.1,1.5);
\node at (-1.3,-0.7) (a1) {$\ket{\psi_i}$};
\node at (1.3,0.7) (a1) {$B$};
\end{tikzpicture}
\end{center}
\caption{\small Two disjoint systems $A$ and $B$ interacting with a large environment $E$. Ontic states of $A$ and $B$ are $\ket{\psi_i}$ and $\ket{\phi_a}$, respectively, However, ontic states of $A+B$ are not generally equal to tensor products $\ket{\psi_i}\otimes\ket{\phi_a}$ and the ontology is ``quantum".}
\label{f2}
\end{figure}

The only information we can have on disjoint sub-systems $A$ and $B$ are the epistemic and ontic states of the combined system $A+B$. The ontic states of the latter will not generally be related to products of ontic states of $A$ and $B$: see figure \ref{f2}. 
This means that it is not generally consistent to make joint ontic assignments and define joint probabilities to sub-systems $A$ and $B$. However, in section \ref{s1.4}, we will see that under suitable circumstances such joint assignment and probabilities can emerge when $A$ and $B$ are macro-systems embedded in a much larger environment.

In addition, as we saw in the last section, the ontic state $\ket{\psi_i}$ of a sub-system $A$  is precisely correlated with the mirror ontic state $\ket{\tilde\psi_i}$ of the $E$, the complement of $A$ in the sense that $A+E$ is the total system. However, in this case the pair of ontic states have no precise relation to that of $A+E$ which is the pure state $\ket{\Psi}$ in \eqref{m34}.

\subsection{Sub-systems and Effective Theories}\label{s1.2}

Since each sub-system $\BA\subset\BS$ enjoys its own set of ontic states and, if the latter are to represent a property assignment, how are we to understand deformations in the definition of the sub-system? This involves discrete changes when we decide to move degrees-of-freedom from the environment $E$ to $\BA$, and vice-versa. The new interpretation  must explain how a recognisably stable classical ontology can emerge out of the myriad of different possible sub-systems and their associated ontic states. 
Essentially it does this by recognising that classical states are not 
directly identified with a particular ontic state of a particular sub-system but rather to coarse grained histories of ontic states and these are not sensitive to the precise microscopic definition of $\BA$. We may change the definition of $A$ by moving microscopic degrees-of-freedom into and out of $A$ without affecting the collective behaviour of $A$ that arises from coarse-grained time averages of a time sequence of ontic states. In fact, we will see that for the sub-system $A$ that is in equilibrium with the environment $E$, its ``classical state" is identified with the ensemble or epistemic state $\hat\rho_A(t)$. The collective dynamics defined by the ensemble is expected to be stable with respect to microscopic re-definitions of $A$.

Before we describe time dependence within the new interpretation, it is important that we establish the limitations of a particular quantum mechanical description of a physical system. Many of the problems with existing modal interpretations result from making unrealistic assumptions about the range of validity of the quantum description. Analysing a quantum system involves 
identifying an appropriate Hilbert space and Hamiltonian such that the resulting dynamics is unitary. However, such descriptions can only be approximately valid above some particular
length or time scale.\footnote{This point was made in \cite{Hollowood:2013cbr} and elaborated in discussions with Jacob Barandes \cite{JB}.} Equivalently, using the uncertainty principle, below a particular momentum or energy scale. For instance, consider a scalar particle. At low enough momenta, non-relativistic quantum mechanics is a good approximation and particle number is effectively a conserved quantity so it make sense to write down effective theories by taking a sector of the full Hilbert space with one particle $\BH_1$. Such a description, however, will break down when the momentum increases and relativistic effects become important. This is governed by a momentum scale $mc$, a length scale $\hbar/mc$ (the Compton wavelength), an energy scale $mc^2$ and a temporal scale $\hbar/mc^2$. In fact, non-relativistic quantum mechanics based on the one-particle sector leads to violations of causality for measurements based on finite spatial regions on the scale $\hbar/mc$. 
In that case, to recover a consistent and causal description one must allow particles to be created and destroyed and the  effective theory in the one-particle sector is no longer valid; one must instead work with a much larger Hilbert space containing all-particle sectors $\BH_0\oplus\BH_1\oplus\cdots$ in order to have a unitary description. Of course, this is where quantum field theory becomes the more appropriate formulation.
At even higher momenta, this description could break down, for instance, if the particle were a composite. At some high momentum scale the constituents would become important and a different effective theory would be needed.

The message here is this that, when we analyse a typical quantum system, we are inevitably doing a low momentum (or low energy, large distance/time scale) approximation. In this effective description there is no sense in which the effective Hilbert space $\BH_\BA$ is a factor of the total Hilbert space of the universe---assuming that the latter even makes sense. In other words, it is not even clear that $\BA$ has a well-defined parent system in the sense that $\BH_\BS=\BH_\BA\otimes\BH_E$. However, in the spirit of effective theory
one can imagine that we can identify $E$ with all the degrees-of-freedom at the scale of the effective theory that directly interact, or are entangled, with $\BA$. The only requirement is that 
$\BS$ is chosen to be large enough in order to achieve an approximately unitary description of the overall dynamics. For overall consistency, it is important that the detailed nature of $E$, the so-called ``environment" or ``bath", is largely irrelevant for the behaviour of the sub-system $\BA$. This turns out to be the case as long as $d_E\gg d_A$. One could say that it is crucial that the environment is present but that its details are largely irrelevant, including the overall state of $\BS$, except in very special situations where the initial state is very non-generic. The latter occurs when a carefully designed measuring device is interacting with a microscopic quantum system.

Given that the analyses of quantum systems are only effective descriptions valid above some particular length/time scale, the so-called ultra-violet cut off, it is important that that the new interpretation 
yields a formalism that that is not 
sensitive to phenomena on the cut off scale. This does not means that the new interpretation  is not applicable to shorter distances or times, but in order to be valid at more refined scales one would need to apply it to the more fundamental effective theory that takes over at this scale. The importance of this observation is that many of the problems suffered by other modal interpretations result from issues that involve arbitrarily short distance and time scales well beyond the validity of the effective theory. The new interpretation, on the contrary, is immune from these difficulties because it acknowledges the inherent limitations of an effective description.
The fact that analyses of quantum systems are only effective, means that there are 
intrinsic errors to any calculation which involve powers of the characteristic length/time scale to the length/time ultra-violet cut off. 

\subsection{Ontic Dynamics}\label{s1.3}

If the issue of how to define ontic states at a given instant of time is simple to state, the issue of how ontic states change in time is a subject fraught with problems for all modal interpretations. 
Our conclusion is that previous approaches are fundamentally flawed because they make unrealistic expectations as to the limits of the validity of the analysis. They do not 
recognise the key fact, discussed in the last section, that an analysis of a quantum system is generally only an effective one and so is valid only on distance and time scales that are greater than some specific ultra-violet cut off. It is then perhaps no surprise that problems arise when one tries to construct a theory of ontic dynamics that is continuous in time. 

A key principle, therefore, is that
it only makes sense to define ontic dynamics on a temporal coarse graining scale that is the ultra-violet cut off scale $\eta$. So for a particle, for instance, this would be $\eta\sim \hbar/mc^2$. As long as this scale is smaller than any characteristic decoherence
time scale of the system under study then the effective description is valid. We denote the
\EQ{
\boxed{\text{characteristic decoherence time scale}=\tau }\nonumber
}
Then the validity of the effective theory requires that
\EQ{
\eta\ll\tau \ .
\label{ap5}
} 
The scale $\tau$ will be defined more precisely later. 
The condition \eqref{ap5} ensures that (i) the effective theory is a valid description and (ii) the coarse graining appears smooth at the scale of time-dependent phenomena of the system.
To summarise, our coarse graining scale lies at ultra-violet cut off scale of the effective theory and for consistency of the effective description, as long as \eqref{ap5} is satisfied and a Markov condition respected, the whole formalism is then insensitive to the exact value of $\eta$ since the discretization errors are of order $\eta/\tau $. However, we cannot attempt to take the cut off $\eta\to0$ and remain within the validity of the effective theory. Note that even the dynamics of the epistemic state governed by the Schr\"odinger equation cannot be considered more fundamental because it is also only an equation valid within the effective theory.

Since it is the interaction between $A$ and $E$ that is responsible for the time dependence  of the probabilities $p_i(t)$, on the time scale $\tau$, and the reason that the ontic states $\ket{\psi_i(t)}$ do not satisfy the Schr\"odinger equation, the fact that the coarse graining scale $\eta\ll\tau $ means that over a time step $\eta$ the set of probabilities and ontic states do not change much. The implication is that there exists a unique one-to-one mapping between ontic states at time $t$ with those at $t+\eta$. We can use the freedom to permute the labels of the ontic states at each time step to ensure that the mapping associates $\ket{\psi_i(t)}$ with $\ket{\psi_i(t+\eta)}$ in the sense that
\EQ{
\bra{\psi_j(t+\eta)}\psi_i(t)\rangle\thicksim\delta_{ij}+{\cal O}(\eta/\tau )\ ,
\label{yu2}
}
in which case, the associated individual probabilities only change by a small amount
\EQ{
p_i(t+\eta)\thicksim p_i(t)\big(1+{\cal O}(\eta/\tau )\big)\ .
}
It is important to emphasise that this continuity condition \eqref{yu2} is defined at the temporal scale $\eta$ and we are not at liberty to take the limit $\eta\to0$ since this would go beyond the domain of applicability of the effective theory. This is fortunate because it saves the new interpretation  from the scourge of macro-flips, a disease that infects other modal interpretations. 
A macro-flip occurs when the eigenvalues $p_i(t)$ and $p_j(t)$, for two macroscopically distinct ontic states, try to cross. But generically the eigenvalues ``repel" each other and this leads to an extremely rapid flip, over a time scale of order $\tau\Delta$, of the ontic state from $\ket{\psi_i}$ to the macroscopically distinct state $\ket{\psi_j}$, or vice-versa.
Given the importance of this issue, we describe it more fully in section \ref{a1}. Having said that, apart from solving the problem of macro-flips, having a small but finite cut off $\eta$ will not affect the dynamics over the physically relevant time scale $\tau$
since the discretisation errors involve powers of $\eta/\tau\ll1$.

Given that the total system $S=A+E$ is assumed to be a pure state $\ket{\Psi(t)}$, the time dependence of the epistemic state $\hat\rho_A(t)$ is determined by solving the Schr\"odinger equation for $S$ giving
\EQ{
\hat\rho_A(t)=\Tr_E\left[\hat U(t,t_0)\ket{\Psi(t_0)}\bra{\Psi(t_0)}\hat U(t_0,t)\right]\ .
\label{shr}
}
where $\hat U(t,t_0)$ is the unitary time evolution operator in $S$.
We want to emphasize that it is really only meaningful to describe this epistemic dynamics on the coarse graining scale $\eta$.

The problem before us is to write down a similar dynamical equation for the probability that the
system $A$ is in the ontic state $\ket{\psi_i(t+\eta)}$ given that it was in the ontic state $\ket{\psi_j(t)}$:
\EQ{
p_{i|j}(t)\equiv p_{i|j}(t+\eta,t)\ .
\label{bst}
}
These conditional probabilities, along with the Markov property discussed below, define a stochastic process. In order to be consistent with the probability constraint \eqref{pc6}, we must have
\EQ{
p_i(t+\eta)=\sum_jp_{i|j}(t)p_j(t)\ ,
\label{pc7}
}
but this does not determine the stochastic process uniquely so its definition is a hypothesis. However, we will argue there are some additional natural constraints which lead us to a unique process:
\begin{description}
\item[Markov] ontic states carry no memory of their previous history and so the conditional probability to be in the ontic state $\ket{\psi_i(t+\eta)}$ should only depends on the ontic state $\ket{\psi_j(t)}$ and not on ontic states at earlier times. This condition is fundamental to our whole approach since it ensures that we can build up the more coarse grained dynamics in terms of the microscopic transitions \eqref{bst} at the ultra-violet scale $\eta$.
\item[Locality] the stochastic process should be driven by the local 
interaction between $A$ and $E$ and so 
in order that it leads to a local form of dynamics we require that it depends only on the 
initial and final ontic states $\ket{\psi_j(t)}$ and $\ket{\psi_i(t+\eta)}$ of $A$ and their ontic mirrors $\ket{\tilde\psi_j(t)}$ and $\ket{\tilde\psi_i(t+\eta)}$ of $E$ as well as $\hat H_\text{int}$ the part of the total Hamiltonian 
\EQ{
\hat H=\hat H_A\otimes I_E+I_A\otimes\hat H_E+\hat H_\text{int}\ ,
\label{b32}
}
that describes the local coupling between $A$ and $E$.\footnote{Issues involving locality should properly be formulated in terms of relativistic quantum field theory, so our notion here is more primitive.} \item[Ergodicity] generically we require that any state can reach any other state in a finite number of steps. However, this must break down for two macroscopically distinct states $\ket{\psi_i}$ and $\ket{\psi_j}$, for which we require that there should only be a minute probability of order $\Delta$ (a typical inner product of two macroscopically distinct states described in section \ref{s1}) of a transition between them over a time scale $\tau$.
\end{description}
As long as the stochastic process has these properties, then its actual microscopic details are largely irrelevant to the behaviour of macro-systems.
The Markov property is very natural given that 
in the new interpretation  the ontic state of a system is just a property assignment at a particular time and has no memory of previous ontic states. Moreover imposing this condition is fundamental because it means that the ultra-violet dynamics $p_{i|j}(t)$ determines the whole stochastic process since over a series of time steps $t_n=t+(n-1)\eta$, according to the Markov property,
\EQ{
p_{j_N|j_1}(t_N,t_1)=\sum_{j_2,\ldots,j_{N-1}}\left[\prod_{n=1}^{N-1}p_{j_{n+1}|j_n}(t_n)\right]\ .
\label{s99}
}
The resulting stochastic process is therefore a conceptually simple discrete-time Markov chain. 

We now turn to the definition of the ultra-violet dynamics.
A useful observation involves the matrix elements
\EQ{
&V_{ij}(t)=\big(p_i(t+\eta)p_j(t)\big)^{1/2}\\ &\times\RE\,\bra{\psi_i(t+\eta)}\otimes\bra{\tilde\psi_i(t+\eta)}\hat U(t+\eta,t)\ket{\psi_j(t)}\otimes\ket{\tilde\psi_j(t)}\ ,
\label{jxx}
}
where $\ket{\tilde\psi_i}$ are mirror ontic states of $E$ defined in \eqref{m34} and $\hat U(t+\eta,t)$ is the unitary time evolution operator in $A+E$. Note that these matrix elements are completely symmetrical between $A$ and $E$ and the resulting dynamics will consequently respect the exact correlation between their ontic states.
One finds from this definition and \eqref{m34}, that
\EQ{
p_i(t+\eta)=\sum_jV_{ij}(t)\ ,\qquad
p_j(t)=\sum_iV_{ij}(t)\ ,
\label{pp2}
}
so it it tempting to relate
\EQ{
V_{ij}(t)\leftrightsquigarrow p_{i|j}(t)p_j(t)\ .
}
But we cannot have equality here, because the matrix elements $V_{ij}(t)$ are not necessarily valued between 0 and 1. However, we can proceed as follows.
If we have labelled the states to be consist with the continuity condition \eqref{yu2}, it follows that since 
\EQ{
p_i(t+\eta)-p_i(t)=\sum_j\big[V_{ij}(t)-V_{ji}(t)\big]\ ,
\label{bx3}
}
the right-hand side must be small of order $\eta/\tau $.
We can define the ultra-violet process by taking, for $i\neq j$,\footnote{Using these expressions it is easy to see that the process defined in \eqref{sp1} and \eqref{sp2} above agrees with the one defined in \cite{Hollowood:2013cbr}: to compare formulae the quantities $p_{ij}^{(n)}$ in \cite{Hollowood:2013cbr} are the conditional probabilities $p_{i|j}(t_n)$ here.}
\EQ{
p_{i|j}(t)=\frac1{p_j(t)}\text{max}\,\Big[V_{ij}(t)-V_{ji}(t),0\Big]
\label{sp1}
}
and
\EQ{
p_{i|i}(t)=1-\sum_{j\neq i}p_{j|i}(t)\ .
\label{sp2}
}
This process satisfies the probability constant \eqref{pc7}.

It is important to realize that there is no guarantee that the process defined above is consistent in the sense that the conditional probabilities $p_{i|j}(t)$ are valued in the interval $[0,1]$. 
In fact, the consistency conditions are
\EQ{
\sum_{j\neq i}p_{j|i}(t)\leq1\ ,\qquad\forall i\ .
\label{cc34}
}
As $\eta\to0$, the elements $p_{i|j}(t)$, $i\neq j$, can be made arbitrarily small and so the process can always be made consistent in this limit. Hence, there is an upper bound on how big the cut off $\eta$ can be taken. In order to investigate the this,
we can interpret $\delta p_i$ for a single time step as being due to a mismatch between the flows into and out of the $i^\text{th}$ state, that is
\EQ{
\delta p_i=\sum_{j\neq i}{p_{i|j}p_j}-\sum_{j\neq i}p_{j|i}p_i\ .
\label{k92}
}
The net flow out of the $i^\text{th}$ state involves the sum on the left-hand side of \eqref{cc34}. If we define the decoherence time scale $\tau$ by
\EQ{
\tau= \eta\Big[\underset{i}\SUP\ \sum_{j\neq i}p_{j|i}\Big]^{-1}\ .
\label{dd45}
}
So consistency of the process requires $\eta\ll\tau$. Note that if $\eta\not\ll\tau$ then this does not imply a breakdown of the formalism but rather a breakdown of the validity of the effective theory: one should go to a more fundamental effective theory valid at smaller distance/time scales.

Generically, both terms on the right-hand side of \eqref{k92} will be of the same order, so that
\EQ{
\delta p_i\thicksim {\cal O}\big(p_i\eta/\tau\big)\ .
\label{k54}
}
However, later, we will describe the situation when $A$ is in equilibrium with the environment, in that case the $p_i$ will be approximately constant on account of a balance between the flows into and out of the states in \eqref{k92}. It is important to notice that ontic states with very small probabilities $p_j(t)$ do not give anomalously large values of $p_{i|j}(t)$ as might be inferred from \eqref{sp1} because the factor of $p_j(t)$ in the denominator is generally balanced by a factor of a similar order in the numerator.

In the limit $\eta\ll\tau$, we can evaluate the matrix elements \eqref{jxx} in perturbation theory:
\EQ{
V_{ij}=\frac{\eta\sqrt{p_ip_j}}\hbar\IM\,\bra{\psi_i}\otimes\bra{\tilde\psi_i}\hat H_\text{int}\ket{\psi_j}\otimes\ket{\tilde\psi_j}+\cdots\ .
\label{frr}
}
The fact that the matrix elements in \eqref{frr} only depends on the coupling $\hat H_\text{int}$ and the ontic states $\ket{\psi_i}$ and $\ket{\psi_j}$ and their mirrors $\ket{\tilde\psi_i}$ and $\ket{\tilde\psi_j}$ encapsulates the locality requirement. Note that at this leading order $V_{ij}=-V_{ji}$.

In addition, the fact that the transition probabilities depend on 
a matrix elements \eqref{frr} involving the states $\ket{\psi_i}\otimes\ket{\tilde\psi_i}$ means that 
the stochastic process will be seen to satisfy the ergodicity requirement. Essentially, if the two states are macroscopically distinct states, then we can expect
$p_{i|j}(t)$ will be suppressed by a factor of order $\Delta$ relative to the generic situation. The probability that the system will make a transition between the ontic states will therefore be vanishing small. In fact, over a time $T$ the chance that the system will make a transition from one macroscopically distinct state to another would be order $T\Delta/\tau$. It is clear, given the crude estimate of
$\Delta$ in section \ref{s1}, we would have to wait of the order of $e^{10^{32}}$ times the age of the universe to see such a transition. 

It is worth emphasising that the ultra-violet dynamics we have defined is not unique. 
The ambiguity corresponds to changing
\EQ{
\delta p_{i|j}=\frac{\Theta_{ij}}{p_j}\ ,\quad
\sum_i\Theta_{ij}=\sum_j\Theta_{ij}=0\ ,
}
subject to the constraint \eqref{cc34}.
However, there are no obvious quantities $\Theta_{ij}$ that could be defined that are at the same time consistent with the ergodicity requirement. 
We take the process that we have defined as being a hypothesis on the same level as the Schr\"odinger equation that determines the dynamics of epistemic state. However, it is possible to take an agnostic point-of-view and avoid a concrete  microscopic definition of the stochastic process because:
\BOX{
As long as the ergodicity condition is satisfied, along with the key probability relation \eqref{pc6}, 
the microscopic details of the stochastic process are actually irrelevant for reproducing standard Copenhagen interpretation phenomenology of macro-systems which are in equilibrium with their environment (as described in section \ref{s2.9}).}

Finally, it is worth making clear the point that, although we have introduced an auxiliary stochastic process to define the dynamics of ontic states, this is completely different from dynamical collapse models discussed in the literature; for example in the review \cite{BG}. 
The latter involve stochastic modifications of Schr\"odinger's equation itself, in other words they involve introducing stochastic dynamics for the epistemic state which is a completely different philosophy from the one we are setting out here.

\subsection{The Continuum Process and Macro-Flips}\label{a1}

It is tempting, even though it runs counter to the methodology of effective theory, to take the stochastic process that we defined in the last section and take the cut off $\eta\to0$, in order to define a continuum process. 
However, typically one should expects pathologies to arise when effective theories are pushed beyond their range of validity. Indeed, in the present case, a pathology manifests as 
the existence of micro-flips that represent a severe problem for existing modal interpretations that insist on following ontic states continuously in time. In this section, we will describe how they arise and how the new interpretation  avoids them.

The ontic states of a sub-system $\ket{\psi_i(t)}$ are defined continuously in time and so it is tempting to define ontic dynamics that is also continuous in time. In fact, the continuum limit of \eqref{bx3} takes the form
\EQ{
\frac{dp_i}{dt}=\sum_jJ_{ij}\ ,\qquad J_{ij}=-J_{ji}\ ,
\label{ge1}
}
with
\EQ{
J_{ij}=\frac{2\sqrt{p_ip_j}}\hbar\,\IM\,\bra{\psi_i}\otimes\bra{\tilde\psi_i}\hat H_\text{int}\ket{\psi_j}\otimes\ket{\tilde\psi_j}\ ,
\label{eqj}
}
which is equal to the perturbative form of $(V_{ij}-V_{ji})/\eta$ using \eqref{frr}.
Written in this form, manifests the fact that if $A$ does not interact with $E$ then  $J_{ij}$ vanishes.\footnote{Note that the expression for $J_{ij}$ seems to be missing a term involving a time derivative compared with \cite{Hollowood:2013cbr,BacciagaluppiDickson:1999dmi}. However, it is easy to see that, since $\partial_t(\ket{\psi_j}\otimes\ket{\tilde\psi_j})=\partial_t\ket{\psi_j}\otimes\ket{\tilde\psi_j}+\ket{\psi_j}\otimes\partial_t\ket{\tilde\psi_j}$ and the sets of states $\{\ket{\psi_j}\}$ and $\{\ket{\tilde\psi_j}\}$ are orthonormal, this term actually vanishes. In addition, the total Hamiltonian may be replaced by $\hat H_\text{int}$ for the same reason.}
In that case, a continuum Markov process can be defined whose master equation takes the form
\EQ{
\frac{dp_i}{dt}=\sum_{j\neq i}\Big(T_{ij}p_j-T_{ji}p_i\Big)\ ,
\label{ge2}
}
corresponding to transitions into and out of  $\ket{\psi_i}$. For the stochastic process satisfying the ergodicity condition defined in the last section
\EQ{
T_{ij}=\frac1{p_j}\,\text{max}\left(J_{ij},0\right)\ .
\label{trt}
}
These are the transition probabilities originally suggested by Bell \cite{Bell:2004suqm} and further analysed in the context of modal interpretations by Bacciagaluppi and Dickson \cite{BacciagaluppiDickson:1999dmi}. This is the continuum version of \eqref{sp1} obtained in the limit $\eta\to0$. 

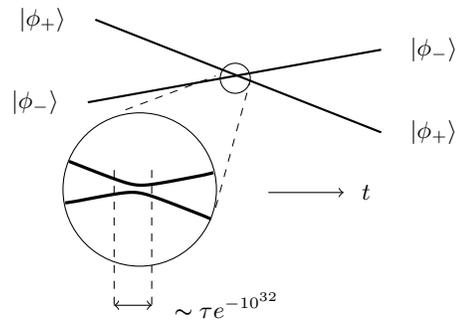
\begin{figure}[ht]
\begin{center}
\begin{tikzpicture}
\draw[very thin,dashed] (1.8,-1.7) -- (2.25,-0.05);
\draw[very thin,dashed] (0.6,-0.4) -- (1.8,0.05);
\begin{scope}[scale=0.2]
\draw (10.3,0.1) circle (1cm);
\draw[-,thick] (1,4) -- (20,-3.5);
\draw[-,thick] (0.5,-1.5) -- (20,2);
\node at (-2.5,4) (a1) {$\ket{\phi_+}$};
\node at (-3,-1.5) (a1) {$\ket{\phi_-}$}; 
\node at (23.5,-3.5) (a1) {$\ket{\phi_+}$};
\node at (23.5,2) (a1) {$\ket{\phi_-}$}; 
\end{scope}
\begin{scope}[scale=0.1,xshift=-2.5cm,yshift=-15cm]
\draw (10.4,0.4) circle (10.2cm);
\draw[-,very thick] plot[smooth] coordinates {(1., 4.14591) (2., 3.75153) (3., 3.3587) (4., 2.96815) (5., 
  2.58114) (6., 2.2) (7., 1.82956) (8., 1.48102) (9., 
  1.1831) (10., 1.) (11., 0.983095) (12., 1.08102) (13., 
  1.22956) (14., 1.4) (15., 1.58114) (16., 1.76815) (17., 
  1.9587) (18., 2.15153) (19., 2.34591) (20.19, 2.64138)};
\draw[-,very thick] plot[smooth] coordinates  {(0.5, -1.38591) (2., -1.15153) (3., -0.958703) (4., -0.768154)
(5., -0.581139) (6., -0.4) (7., -0.229563) (8., -0.081025) (9., 
  0.0169048) (10.,0.) (11., -0.183095) (12., -0.481025) (13., -0.829563) (14.,
-1.2) (15., -1.58114) (16., -1.96815) (17., -2.3587) (18.,-2.75153) (19., -3.14591) (19.9, -3.54138)};
\draw[thin,dashed] (7,3) -- (7,-15);
\draw[thin,dashed] (12,3) -- (12,-15);
\draw[<->,thin] (7,-15) -- (12,-15);
\node at (22,-15) (a1) {$\thicksim \tau  e^{-10^{32}}$};
\end{scope}
\draw[->] (2.5,-1.5) -- (3.5,-1.5);
\node at (3.8,-1.5) (b1) {$t$};
\end{tikzpicture}
\end{center}
\caption{\small A crossover where there is an approximate degeneracy as $p_+(t)$ and $p_-(t)$ try to cross but ultimately ``repel" on a time scale of order $\tau \Delta$. Any continuous-time stochastic process will involve a rapid switching between macroscopically distinct states $\ket{\phi_\pm}$---a macro-flip. On the contrary, our coarse-grained stochastic process cannot ``resolve" the crossover and a macro-flip does not occur even though the cut-off scale $\eta$ is much smaller than any other characteristic time scales in the problem: $\tau \gg\eta\gg\tau \Delta$.}
\label{f6}
\end{figure}

Now we turn to the issue of macro-flips.
The problem of the continuous stochastic process above, occurs when two eigenvalues of $\hat\rho_A(t)$, say $p_+(t)$ and $p_-(t)$, associated to macroscopically distinct ontic states, try to cross as illustrated in \ref{f6}. 

If the eigenvalues actually \emph{do\/} cross then there is no problem because then one can define the continuity of the ontic states in time by imposing analyticity across the point of degeneracy. This ensures that there are no abrupt transitions between the two macroscopically distinct states.
However, generically the eigenvalues will not cross and, on the contrary, there will be a crossover. In order to analyse what happens we can  
isolate the important part of the reduced density matrix in the 2-dimensional subspace spanned by the exactly orthogonal macroscopically distinct states $\ket{\phi_\pm}$ roughly constant in the neighbourhood of  the degenerate point at $t=0$:
\EQ{
\rho_\text{subspace}\thicksim\left(\begin{array}{cc} p_0+a_1t& p_0\Delta\\ p_0\Delta& p_0+a_2t\end{array}\right)\ .
}
Here, $a_i$ and $p_0$ are real constants and we define $a=(a_1-a_2)/2$. Note that the $a_i$ will be of the order $\tau ^{-1}$, where $\tau $ is, as previously, a characteristic decoherence scale. Here, $\Delta$ is a measure of the typical inner-product of two macroscopically distinct states. As described in section \ref{s1}, for a typical macroscopic system we might have $\Delta\sim e^{-10^{32}}$.
In the absence of the tiny inner product between the macroscopically distinct states, i.e.~when $\Delta=0$, the ontic states (eigenvectors) are $\ket{\psi_i}=\ket{\phi_i}$ and the eigenvalues actually cross at $t=0$. However, when $\Delta\neq0$ the eigenvalues are
\EQ{
p_\pm(t)=p_0+\frac{a_1+a_2}2t\pm\sqrt{(at)^2+(p_0\Delta)^2}
}
and the level crossing is avoided since, although the two eigenvalues become close, they 
never actually cross. Near the crossover, the eigenvectors are approximately
\EQ{
\ket{\psi_\pm(t)}=\cos\theta(t)\ket{\phi_\pm}\pm\sin\theta(t)\ket{\phi_\mp}\ ,}
where
\EQ{
\tan\theta(t)=\frac{at+\sqrt{(at)^2+(p_0\Delta)^2}}{p_0\Delta}\ ,
}
which goes from $0$ to $\frac\pi2$ as $t$ increases through the crossover. The time for the crossover to occur is order $\tau \Delta$.
This reveals the problem: by avoiding the degeneracy the system exhibits an instability in the sense that it flips between the macroscopically distinct states $\ket{\phi_+}$ and $\ket{\phi_-}$ in a very short time of order $\tau \Delta$. This time scale, for any system, will be much smaller than even the Planck scale and therefore certainly much smaller than the cut off scale $\eta$.
So an interpretation based on the continuous time stochastic process suffers from unacceptably rapid switching between macroscopically distinct states---a macro-flip---whenever probabilities try to cross. Of course such macro-flips are completely unphysical and one can only conclude that the continuous-time stochastic process, pushing as it does beyond the domain of applicability of the effective theory, is fundamentally flawed.

On the contrary, our coarse-grained stochastic process defined in section \ref{s1.3} is immune from macro-flips. The reason is that the crossover scale $\tau \Delta\ll\eta$, even though $\eta\ll\tau $, and so a crossover simply cannot be resolved by the coarse-grained stochastic process. The smoothness condition \eqref{yu2} then ensures that across the time step that includes the crossover the ontic states are preserved $\ket{\psi_i(t+\eta)}\approx\ket{\psi_i(t)}\approx\ket{\phi_i}$, to order $\eta/\tau $, and no dramatic macro-flip of the ontic state occurs.

\section{The Classical Limit}\label{s2}

\subsection{Emergent Classicality}\label{s1.4}

It is key feature of the new interpretation  that what we ordinarily understand as classical behaviour only emerges in appropriate situations and is not put in by hand.
The fact that a successful phenomenology of macro-systems arises  rests largely on the fact that the stochastic process defined in the last section satisfies the ergodicity condition. 

More fundamentally, the new interpretation makes statements about ontology that are precise but they are not necessarily of a classical kind. For instance, as discussed in section \ref{s1.12}, an interpretation based on reduced density matrices cannot generally make joint property assignments to two disjoint systems $A$ and $B$ let alone associate probabilities to them.
The only statements that can be made involving both $A$ and $B$ are via the combined system $A+B$ and in this case
its ontic states will generally not be a tensor product of ontic states of the sub-systems, i.e.~$\ket{\psi_i}\otimes\ket{\phi_a}$. The relation between ontic states of the three systems $A$, $B$ and $A+B$ will be more fuzzy and potentially contradictory. One could say that the fragments of reality cannot be drawn together to form a consistent whole. We call this a ``quantum" ontology. 

However, a familiar ``classical" ontology can be an emergent phenomenon in the following sense. 
Suppose $A$ and $B$ are two disjoint weakly-interacting or causally separated macro-systems in, or close to, equilibrium and hence strongly entangled with the environment $E$ with $d_A,d_B\ll d_E$. We expect in these circumstances that the ontic states of $A+B$ will indeed approximately factorize into a tensor product of the ontic states of $A$ and $B$. The mismatch will involve typically minute order $\Delta$ effects:
\EQ{
\ket{\Phi_{m(i,a)}}=\ket{\psi_i}\otimes\ket{\phi_a}+{\cal O}(\Delta)\ ,
\label{mdr}
}
where $m=m(i,a)$ is a 1-to-1 map. What this means is the descriptions provided by $A$, $B$ and $A+B$ can be integrated into a consistent whole, at least to high degree of accuracy.  

In this context, it is meaningful to make joint property assignments for $A$ and $B$ and 
we can interpret, in an emergent sense, the probability $p_{m(i,a)}$ as the joint probability for a pair of ontic states $\ket{\psi_i}$ and $\ket{\phi_a}$ of $A$ and $B$:
\EQ{
p(i,a)\overset{\text{emergent}}=p_{m(i,a)}\ .
\label{jpb}
}
It is the view from $A+B$ that is needed to follow any potential correlations between the ontic states of $A$ and $B$. Note also that the ontic states of $A$ or $B$ still cannot generally be taken as global property assignments outside of the triplet of sub-systems $A$, $B$ and $A+B$.

The emergent joint probabilities satisfy the usual probability relations, but only to order $\Delta$,
\EQ{
&\sum_a p(i,a)=p_i+{\cal O}(\Delta)\ ,\\&\sum_ip(i,a)=p_a+{\cal O}(\Delta)\ ,
}
although $\sum_{ia}p(i,a)=1$.

In general, the ontic states of $A$ and $B$ can be correlated in a classical (non-entangled) sense when $\hat\rho_{A+B}\neq\hat\rho_A\otimes\hat\rho_B$ meaning that
\EQ{
p(i,a)\neq p_i\,p_a\ .
}
The case when $A$ and $B$ are not correlated corresponds to when $\hat\rho_{A+B}=\hat\rho_A\otimes\hat\rho_B$ and $p(i,a)=p_ip_a$. 

Note that we will meet an example in section \ref{s4.1} an example where the ontic states of $A+B$ are tensor product states of $A$ and $B$ but the factors of one of them are not the ontic states of the corresponding sub-system. So simply being a tensor product state is not sufficient to have a ``classical" ontology.

The picture above of emergent joint ontic assignments and joint probabilities can be generalized to many weakly interacting or causally disconnected macro-systems $A_1+\cdots+A_n$. An ontic state of the parent system will be, to order $\Delta$, a tensor product state $\ket{\Phi_{m(i^{(1)},\ldots,i^{(n)})}}=\ket{\psi_{i^{(1)}}^{(1)}}\otimes\cdots\otimes\ket{\psi_{i^{(n)}}^{(n)}}$, and so emergent joint probabilities can be defined of the form
\EQ{
p\big(i^{(1)},\ldots,i^{(n)}\big)\overset{\text{emergent}}=p_{m(i^{(1)},\ldots,i^{(n)})}\ .
}

The picture we have here is that the emergent classical world involves 
patching together very slightly different descriptions---differing at ${\cal O}(\Delta)$---of the same systems from the point-of-view of $A_i$, $A_i+A_j$, $A_i+A_j+A_k$, etc. 
For more microscopic systems this integration of ontic states becomes more ambiguous and a classical description evaporates to be replaced by an ontology that is truly quantum. 
We can quantify the degree of classicality in terms of the generic scale $\Delta$.
So we can expect systems to exhibit quantum fuzziness when $\Delta$ is not so small so that  the relation between the ontic states becomes ambiguous and joint ontic assignments and joint probabilities cannot be consistently defined.

\subsection{Link with Statistical Mechanics}\label{s2.9}

A key requirement of an interpretation of quantum mechanics is to explain how the classical
behaviour of macroscopic systems emerges. Macro-systems with many degrees-of-freedom are complicated systems whose collective behaviour is captured by the techniques of statistical mechanics. It seems natural that any quantum origin of classical behaviour must, at the very least, be able to give a consistent foundation to classical statistical mechanics. In fact, we might hope that such an understanding would put classical statistical mechanics on a firmer conceptual footing given that it 
is still, somewhat surprisingly, a controversial subject. In particular, there is no consensus on the role of probability, the meaning of entropy and the relation of ensemble averages to time averages. 

In the last few year a rather different and intrinsically quantum approach to the subject has been developed \cite{PopescuShortWinter:2005fsmeisa,PopescuShortWinter:2006efsm,BL,LL,GoldsteinLebowitzTumulkaZanghi:2006ct,LPSW,Sh,GMM}.
In this approach, ensembles arise at the quantum level when a system $\BA$ is entangled with a large thermal bath, or environment, $E$. So the total system $S=\BA+E$ can be in a pure state but, nevertheless, the sub-system $\BA$ has a reduced density matrix that defines an ensemble $\hat\rho_\BA$. In this point-of-view, the thermodynamic entropy of $\BA$ is precisely the entanglement entropy of the sub-system $S=-\text{Tr}(\hat\rho_\BA\log\hat\rho_\BA)$ which is non-vanishing when $\BA$ is non-trivially entangled with the bath.

When the bath is much bigger than the system $d_E\gg d_A$ there are some very powerful principles that emerge.\footnote{The following discussion here is taken mainly from  Popescu, Short and Winter
\cite{PopescuShortWinter:2005fsmeisa} and Linden, Popescu, Short and Winter \cite{LPSW,Sh}.}
For almost any pure state of the total system in some subspace $\BH_R\subset \BH_S$ described by some global constraint $R$ (preserved under time evolution) the reduced density matrix of the system $\BA$ is approximately equal to
\EQ{
\hat\rho_\BA\approx\frac{\text{Tr}_E\hat\PP_R}{d_R}\ ,
\label{dpp}
}
where $\hat\PP_R$ is the projection operator on the subspace $\BH_R$. If the global constraint is on the energy and the interaction between the system $\BA$ and the environment $E$ is sufficiently weak then it is straightforward to show that $\hat\rho_\BA$ is approximately the canonical ensemble,
\EQ{
\hat\rho_\BA\approx\frac{e^{-\beta\hat H_\BA}}Z\ ,\qquad Z=\text{Tr}_A\,e^{-\beta\hat H_\BA}\ .
}
However, it is important that the principle applies to other possible constraints $R$ on the total system, including the absence of a constraint, and also to cases where the interaction between the system and bath is not small. 

The implications for statistical mechanics are evident. If the total system starts out in almost any state in the subspace $\BH_R$, then the state of the sub-system $\BA$ is approximately the state \eqref{dpp} independent of time. This describes a situation where the system is in equilibrium with its environment. In this analysis, the entropy becomes an objective property of the state of $\BA$ caused by entanglement with $E$. 

More general questions involve situations when the system starts off in a state which is not in equilibrium \cite{LPSW,Sh}. The results are summarised below:\footnote{It is also possible to incorporate the constraint $R$ on the system.}
\begin{description}
\item[Equilibration] subject to some reasonable conditions, every pure state of $S$ is such that a small sub-system $A\subset S$ will equilibrate meaning that $\hat\rho_A(t)$ approaches a limit which fluctuates about a constant. Note that the initial state does not need to be ``typical", in other words even though the overwhelming number of states of $S$ are such that $A$ is in equilibrium already, it is also true of initial states where $A$ is far from equilibrium. This includes tensor product
states $\ket{\Psi}=\ket{\psi}\otimes\ket{\phi}$.
\item[Bath independence] in the case that the initial state is a tensor product $\ket{\Psi}=\ket{\psi}\otimes\ket{\phi}$, the equilibrium state of $A$ is independent of the state of the bath $\ket{\phi}$.
\item[Sub-system independence] there are general conditions under which the equilibrium state of $A$ is independent of $\ket{\psi}$ the initial state of $A$. However, there are also non-generic situations for which the equilibrium state depends sensitively on $\ket{\psi}$.
\end{description}
It is clear that the focus on sub-systems means that this re-formulation of statistical mechanics is closely related with modal quantum mechanics. Moreover, the new interpretation  adds a new and important detail to the story through the existence of the ontic state of the sub-system $A$.
This is analogous to the micro-state of classical statistical mechanics. However, it is important to point out that its dynamics, described by the stochastic process described in section \ref{s1.3}, is conceptually simpler than the dynamics of micro-states in classical statistical mechanics because in the quantum case the number of ontic states is always finite and the stochastic process is a simple discrete-time Markov chain.

When the system $A$ equilibrates, $\hat\rho_A(t)$ and its eigenvalues will fluctuate around a slowly varying quasi-equilibrium and the underlying Markov chain becomes approximately {\it homogeneous\/}: that is the transition matrix $p_{i|j}(t)$ becomes time independent over time scales of order $\tau$. Of course, there may be much slower time dependence for on scales $\gg\tau$. 
The rate of flow into and out of each ontic state in \eqref{k92} approximately balances. 
Under generic conditions, although the microscopic transitions between a pair of states only go one way, over a finite number $n$ of time steps $p_{i|j}(t+n\eta,t)$ is a matrix whose entries are all $>0$ and hence the equilibrium process is {\it regular\/}. The meaning of this is that any ontic state is only a finite number of time steps away from any other state. The fact that the process is a {\it regular\/} homogeneous Markov chain implies that it is also {\it ergodic\/} and then it is a standard result that 
\EQ{
\lim_{n\to\infty}p_{i|j}(t+n\eta,t)=p_i(t+n\eta)\ ,
}
independent of $j$, and so, whatever the initial state, after a large number of steps, the probability distribution is equal to $p_i(t)$. More precisely, one can show that the number of time steps must be at least order $\tau/\eta$; in other words, one must wait for a time of the order of $\tau$, the decoherence time defined by \eqref{dd45}, for the memory of the initial state to be lost:
\EQ{
\boxed{\text{equilibrium:}\qquad p_{i|j}(t+\tau,t)\approx p_i(t+\tau)}
}
This means that, in equilibrium, the actual ultra-violet details of the stochastic process are hidden over time scales of order $\tau$, the decoherence time.\footnote{As a simple example of an equilibrium process, suppose that at equilibrium $A$ is maximally entangled with $E$, so $p_i\approx1/d_A$. If we define the process by taking one of each pair $\{p_{i|j},p_{j|i}\}$ randomly and giving it the value $p$, while the other vanishes. In this case, if one takes $p_{i|j}(t+T,t)-p_{i|k}(t+T,t)$ for some randomly chosen distinct $i$, $j$ and $k$, then this approaches zero as $\exp(-T/\tau)$, with the time to approach equilibrium $\tau=2\eta/(pd_A)$, valid when $\tau\gg\eta$. On the other hand, the decoherence time is defined in \eqref{dd45} as $\eta\big[\sum_{j\neq i}p_{j|i}\big]^{-1}\approx 2\eta/(pd_A)=\tau$  so this confirms that the equilibrium time is equal to the decoherence time.}

In addition, ergodicity means that, in equilibrium, the time average of a temporal sequence of ontic states over a time scale of order $\tau$ is well approximated by the ensemble average described by $\hat\rho_A(t)$.
This gives us another way to interpret the single-time probabilities $p_i(t)$: when $A$ has reached equilibrium with the bath, the single-time probabilities $p_i(t)$ are approximately constant and are equal to the probability that, in a suitably coarse-grained time average, the system is in the ontic state $\ket{\psi_i(t)}$, independent of the initial state in the past. From the point-of-view of the emergent classical view, the ontic state of a macro-system in equilibrium is hidden and the probabilities $p_i(t)$ can be given the ignorance interpretation:
\BOX{When a system is in equilibrium with its environment, the ensemble associated to the---approximately constant---reduced density matrix $\hat\rho_A(t)$ captures the time average over the dynamics of the ontic state. So we can associate the classical description of a macro-system with the collective behaviour of the ensemble $\hat\rho_A(t)$.}

\begin{figure}[ht]
\begin{center}
\begin{tikzpicture}[yscale=0.4,xscale=0.4]
\begin{scope}[xshift=-8cm,yshift=-3cm,xscale=0.5,yscale=0.5]
\draw[-] (0.6,0.6) -- (5.4,0.6) -- (5.4,10.4) -- (0.4,10.4) -- (0.6,0.6);
\foreach \position in {(1, 1), (1, 2), (1, 3), (1, 4), (1, 5), (1, 6), (1, 7), (1, 8), (1, 
  9), (1, 10), (2, 1), (2, 2), (2, 3), (2, 4), (2, 5), (2, 6), (2, 
  7), (2, 8), (2, 9), (2, 10), (3, 1), (3, 2), (3, 3), (3, 4), (3, 
  5), (3, 6), (3, 7), (3, 8), (3, 9), (3, 10), (4, 1), (4, 2), (4, 
  3), (4, 4), (4, 5), (4, 6), (4, 7), (4, 8), (4, 9), (4, 10), (5, 
  1), (5, 2), (5, 3), (5, 4), (5, 5), (5, 6), (5, 7), (5, 8), (5, 
  9), (5, 10)}
\filldraw[black] \position circle (0.05cm);
\end{scope}
\begin{scope}[xshift=-4cm,yshift=-3cm,xscale=0.5,yscale=0.5]
\draw[-] (0.6,0.6) -- (5.4,0.6) -- (5.4,10.4) -- (0.4,10.4) -- (0.6,0.6);
\foreach \position in {(1, 1), (1, 2), (1, 3), (1, 4), (1, 5), (1, 6), (1, 7), (1, 8), (1, 
  9), (1, 10), (2, 1), (2, 2), (2, 3), (2, 4), (2, 5), (2, 6), (2, 
  7), (2, 8), (2, 9), (2, 10), (3, 1), (3, 2), (3, 3), (3, 4), (3, 
  5), (3, 6), (3, 7), (3, 8), (3, 9), (3, 10), (4, 1), (4, 2), (4, 
  3), (4, 4), (4, 5), (4, 6), (4, 7), (4, 8), (4, 9), (4, 10), (5, 
  1), (5, 2), (5, 3), (5, 4), (5, 5), (5, 6), (5, 7), (5, 8), (5, 
  9), (5, 10)}
\filldraw[black] \position circle (0.05cm);\end{scope}
\begin{scope}[xshift=0cm,yshift=-4.1cm,xscale=0.5,yscale=0.76]
\draw[-] (0.6,0.7) -- (5.4,0.7) -- (5.4,10.3) -- (0.4,10.3) -- (0.6,0.7);
\draw[densely dashed] (0.6,5.5) -- (5.4,5.5);
\foreach \position in {(1, 1), (1, 2), (1, 3), (1, 4), (1, 5), (1, 6), (1, 7), (1, 8), (1, 
  9), (1, 10), (2, 1), (2, 2), (2, 3), (2, 4), (2, 5), (2, 6), (2, 
  7), (2, 8), (2, 9), (2, 10), (3, 1), (3, 2), (3, 3), (3, 4), (3, 
  5), (3, 6), (3, 7), (3, 8), (3, 9), (3, 10), (4, 1), (4, 2), (4, 
  3), (4, 4), (4, 5), (4, 6), (4, 7), (4, 8), (4, 9), (4, 10), (5, 
  1), (5, 2), (5, 3), (5, 4), (5, 5), (5, 6), (5, 7), (5, 8), (5, 
  9), (5, 10)}
\filldraw[black] \position circle (0.05cm);
\end{scope}
\begin{scope}[xshift=4cm,yshift=2cm,xscale=0.5,yscale=0.5]
\draw[-] (0.6,0.6) -- (5.4,0.6) -- (5.4,5.4) -- (0.4,5.4) -- (0.6,0.6);
\foreach \position in {(1, 1), (1, 2), (1, 3), (1, 4), (1, 5), (2, 1), (2, 2), (2, 3), (2, 4), (2, 5),  (3, 1), (3, 2), (3, 3), (3, 4), (3, 
  5),  (4, 1), (4, 2), (4, 
  3), (4, 4), (4, 5),  (5, 
  1), (5, 2), (5, 3), (5, 4), (5, 5)}
\filldraw[black] \position circle (0.05cm);
\end{scope}
\begin{scope}[xshift=4cm,yshift=-5cm,xscale=0.5,yscale=0.5]
\draw[-] (0.6,0.6) -- (5.4,0.6) -- (5.4,5.4) -- (0.4,5.4) -- (0.6,0.6);
\foreach \position in {(1, 1), (1, 2), (1, 3), (1, 4), (1, 5), (2, 1), (2, 2), (2, 3), (2, 4), (2, 5),  (3, 1), (3, 2), (3, 3), (3, 4), (3, 
  5),  (4, 1), (4, 2), (4, 
  3), (4, 4), (4, 5),  (5, 
  1), (5, 2), (5, 3), (5, 4), (5, 5)}
\filldraw[black] \position circle (0.05cm); 
\end{scope}
\draw[very thick,->]  (-9,0.8) to[out=0,in=180] (-6.7,1);
\draw[very thick,->]  (-6.3,1) to[out=0,in=180] (-2.2,-1.5);
\draw[very thick,->] (-1.8,-1.5) to[out=0,in=180]  (1.8,1.2);
\draw[very thick,->]  (2.2,1.2) to[out=0,in=180] (5.3,3); 
\draw[very thick,->]  (5.7,3) to[out=0,in=180] (8.2,3.7); 
\draw[very thick,->]  (-9,-1.8) to[out=0,in=180] (-6.2,0);
\draw[very thick,->]  (-5.8,0) to[out=0,in=180] (-2.7,1.5);
\draw[very thick,->] (-2.3,1.5) to[out=0,in=180]  (2.3,-1.8);
\draw[very thick,->]  (2.7,-1.8) to[out=0,in=180] (5.8,-3.5); 
\draw[very thick,->]  (6.2,-3.5) to[out=0,in=180] (8.2,-3.7); 
\draw[->] (-5,-5) -- (-1,-5);
\node at (-9,4) (b2) {ontic state};
\node at (-5,5) (b1) {epistemic state $\hat\rho_A$};
\draw[->] (b2) -- (-6.6,2);
\draw[->] (b1) -- (-5.5,2.4);
\node at (-0.2,-5) (a1) {$t$};
\node at (7,0) (b5) {sub-ensembles};
\node at (7.8,1.2) (b3) {$\hat\rho_A^{(1)}$};
\node at (7.8,-1.2) (b4) {$\hat\rho_A^{(2)}$};
\draw[->] (b3) -- (6.5,2.2);
\draw[->] (b4) -- (6.5,-2.2);
 \end{tikzpicture}
    \end{center}
  \caption{\small An illustration of the process of ergodicity breaking during a measurement. Here, the set of of ontic states splits into two distinct sub-ensembles between which the probability of transition is vanishingly small. Two ontic histories are shown that end up in different sub-ensembles after the measurement.}
\label{f8}
\end{figure}
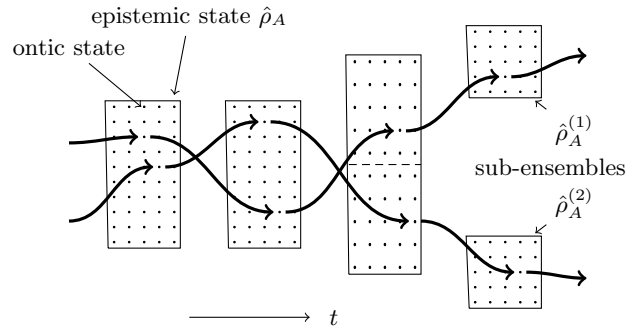

It is noteworthy that the equilibrium state $\hat\rho_A(t)$ does not depend on the details of the initial state of the bath, however, the same cannot be said in all circumstances for the initial state of $A$. In certain situations, discussed in \cite{LPSW}, the final equilibrium epistemic state of $A$ can depend very sensitively on the initial state. This is connected to a breakdown of ergodicity of the stochastic process.
We can expect this to happen when the equilibrium state of $A$ has sets of ontic states which are macroscopically distinct. 
In these conditions, the matrix 
elements $V_{ij}(t)$ between states in different sets will be minute of order $\eta\Delta/\tau$. 
In this case, ergodicity of the stochastic process is broken and once the system has equilibrated the resulting time average of a temporal sequence of ontic states is then only captured by a sub-ensemble of $\hat\rho_\BA(t)$. We could summarise the situation by saying that:
\BOX{In the new interpretation, both the correlata---the ergodic subsets of states described by the sub-ensembles---and the correlations, in the form of the joint probabilities \eqref{jpb}, are emergent quantities.}
In this situation, the dynamics of the underlying ontic state as described by the stochastic process is very sensitive to the initial state of $A$.
This is precisely what happens in a phase transition in a statistical system like the Ising magnetic described in the introduction as the temperature is lowered. But it also describes what happens in a quantum measurement where the measuring device is designed to be very sensitive to the quantum state of the system being measured. We will argue in \S\ref{s3}, that when the measuring device equilibrates with the environment there is a breaking of ergodicity of the underling Markov chain such that each ergodic component corresponds to a distinct measurement outcome. Collapse of the wave function corresponds to a tidying up exercise in which one removes the ergodic components that are not reachable from the component that corresponds to the particular measurement outcome that is realized, as illustrated in figure \ref{f8}.

\section{Measurements and the Collapse of the Wave Function}\label{s3}

In this section we show how the new interpretation  can give a convincing account of the measurement process. There are some benefits in starting with a simple model, showing how it successfully describes certain aspects but also how it has certain limitations. Then we introduce a more sophisticated model which preserves the good features and solves the problems. 
This shows that models need to be realistic otherwise one can be led to misleading conclusions. Along the way, we will need to use intuition from the 
theory of decoherence, e.g.~\cite{Bohm:1951qt,JoosZeh:1985ecptiwe,Joos:2003dacwqt,Zurek:2003dtqc,Schlosshauer:2005dmpiqm,SchlosshauerCamilleri:2008qct,BreuerPetruccione:2002toqs}.

\subsection{A Na\"\i ve Model}\label{s2.1}

The simplest set up involves an idealized von Neumann measurement on a microscopic system. Let us suppose the system $\BP$ is some finite dimensional quantum system with some basis states  $\ket{\xi_i}$ that are eigenstates of the operator associated to the observable we want to measure. The system $\BP$ is then coupled to a measuring device $\BA$ through a Hamiltonian which acts as
\EQ{
\hat H\ket{\xi_i}\otimes\ket{\psi}=\ket{\xi_i}\otimes\hat H^{(i)}\ket{\psi}\ ,
}
If the initial state of the combined system is taken to be the non-entangled state
\EQ{
\ket{\Psi(0)}=\Big[\sum_ic_i\ket{\xi_i}\Big]\otimes\ket{\psi(0)}\ ,
}
then using the linearity of the Schr\"odinger Equation we have
\EQ{
\ket{\Psi(t)}=\sum_ic_i\ket{\xi_i}\otimes\ket{\psi_i(t)}\ ,
\label{b23}
}
where $\ket{\psi_i(t)}$ is the state that evolves from $\ket{\psi(0)}$ via the effective Hamiltonian $\hat H^{(i)}$. In order for $\BA$ to be efficacious, it must be that the states $\ket{\psi_i(t)}$ become macroscopically distinct after a macroscopic time $T$. This is the process of decoherence \cite{Bohm:1951qt,JoosZeh:1985ecptiwe,Joos:2003dacwqt,Zurek:2003dtqc,Schlosshauer:2005dmpiqm,SchlosshauerCamilleri:2008qct,BreuerPetruccione:2002toqs}. We expect the inner product of the states
$\ket{\psi_i(t)}$ to exhibit a behaviour roughly of the form
\EQ{
\big|\bra{\psi_{i}(t)}\psi_{j}(t)\rangle\big|\thicksim \exp\left[-NX(t)^2/\ell^2\right]\ ,
}
for $i\neq j$,
where $X(t)$ describes how the distance between the microscopic constituents of the measuring device behaves between the two measurement outcomes. We expect that this goes from 
$0$ to the macroscopic scale $L$ at the end of the measurement at $t=T$, and so
\EQ{
\big|\bra{\psi_i(T)}\psi_j(T)\rangle\big|\thicksim\Delta\ ,\qquad i\neq j\ .
}
We can estimate how fast decoherence occurs compared with the 
macroscopic measuring time $T$ by assuming that $X(t)$ is linear in $t$. As an example, by taking the same values for $N$, $\ell$ and $L$ as in section \ref{s1}, we find
\EQ{
\frac{\tau }T\thicksim\frac{\ell}{L\sqrt{N}}\leq e^{-16}\ .
}
\begin{figure}[ht]
\begin{center}
\begin{tikzpicture}[yscale=0.15,xscale=0.23]
\draw[->] (0,0) -- (20,0);
\draw[->] (0,0) -- (0,21);
\node at (21,0) (a1) {$t$};
\node at (-1.2,19) (a1) {$1$};
\node at (-1.2,0) (a1) {$0$};
\node at (7,13.2) (b1) {$p_1(t)$};
\node at (11,7.2) (b2) {$p_2(t)$};
\node at (15,3.6) (b3) {$p_3(t)$};
\node at (-4,24) (b4) {$\ket{\psi(0)}$};
\draw[->] (b4) -- (-0.2,19);
\node[rotate=90] at (-2,10) (a1) {probability};
\draw[decoration={brace,amplitude=0.5em},decorate] (8,20) -- (19,20);
\draw[decoration={brace,amplitude=0.5em},decorate] (0.4,20) -- (6.5,20);
\node at (3.3,22.5) (z1) {decoherence};
\node at (13.5,26.5) (a1) {$\ket{\psi_j(t)}$};
\node at (13.5,24.5) (a1) {macroscopically};
\node at (13.5,22.5) (a1) {distinct};
\draw[-] (6.5,0) -- (6.5,-0.5);
\node at (19,-2) (d1) {$T$}; 
\draw[-] (19,0) -- (19,-0.5);
\node at (6.5,-2) (d1) {$\tau $}; \draw[very thick] plot[smooth] coordinates {(0, 19.)  (1, 18.2124)  (2, 16.4584)};
\draw[->,very thick]  (2, 16.4584) --  (2, 2.35062);
\draw[densely dashed] plot[smooth] coordinates {(2, 16.4584)  (3, 14.6642)  (4, 13.2187)  (5,
   12.2299)  (6, 11.6976)  (7, 11.4838)  (8, 11.4189)  (9, 
  11.4035)  (10, 11.4005)  (11, 11.4001)  (12, 11.4)  (13, 11.4)  (14,
   11.4)  (15, 11.4)  (16, 11.4)  (17, 11.4)  (18, 11.4)  (19, 11.4)};
\draw[densely dashed] plot[smooth] coordinates {(0, 0)  (1, 0.773204)  (2, 2.35062) };
\draw[very thick] plot[smooth] coordinates { (2, 2.35062)  (3, 3.65956) (4,4.51829)  (5, 5.12001) };
\draw[->,very thick]  (5, 5.12001)  --   (5, 1.65009) ;
\draw[densely dashed] plot[smooth] coordinates
  {(5, 5.12001)  (6, 5.48233)  (7, 5.6376)  (8, 5.68586)  (9,
   5.69741)  (10, 5.69961)  (11, 5.69995)  (12, 5.7)  (13, 5.7)  (14, 
  5.7)  (15, 5.7)  (16, 5.7)  (17, 5.7)  (18, 5.7)  (19, 5.7)};
\draw[densely dashed] plot[smooth] coordinates  {(0, 0)  (1, 0.0143922)  (2, 0.191004)  (3, 
  0.676266)  (4, 1.26304) (5, 1.65009)};
 \draw[very thick] plot[smooth] coordinates  {  (5, 1.65009)  (6, 1.82005)  (7, 
  1.87861)  (8, 1.89526)  (9, 1.89913)  (10, 1.89987)  (11, 
  1.89998)  (12, 1.9)  (13, 1.9)  (14, 1.9)  (15, 1.9)  (16, 
  1.9)  (17, 1.9)  (18, 1.9)  (19, 1.9)};
\end{tikzpicture}
    \end{center}
  \caption{\small An example of how the probabilities $p_i(t)$ might behave for the case $d_P=3$.  During the decoherence period $0\leq t\leq\tau$, the states
$\ket{\psi_i(t)}$ become macroscopically distinct while the probabilities change in time. At the end of the measurement at $T\gg\tau $, the eigenstates of the reduced density matrix are approximately the states $\ket{\psi_i(T)}$ that occur with probability $p_i(T)$. Also show in bold is a hypothetic solution of the stochastic process in this case with two transitions.}
\label{f5}
\end{figure}
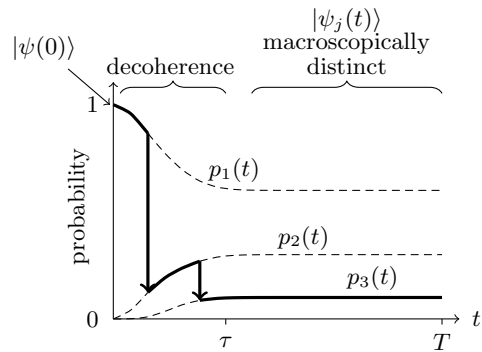

Now we apply the new interpretation  to the model. The reduced density matrix of the measuring device is
\EQ{
\hat\rho_\BA(t)=\sum_i|c_i|^2\ket{\psi_i(t)}\bra{\psi_i(t)}\ .
}
We can now follow the ontic states in time as decoherence occurs \cite{1993PhRvD..48.3768A,Hollowood:2013cbr}: see figure \ref{f5}. In this case, one can verify that only the transition probabilities $p_{2|1}(t)$, $p_{3|1}(t)$ and $p_{3|2}(t)$ are non-vanishing and so the histories of ontic states are rather simple as illustrated in figure \ref{f5}.
After the measurement is complete 
$t=T$, the ontic states of $\BA$ are approximately equal to the $\ket{\psi_i(T)}$ up to terms which are order $\Delta$. But, as emphasised earlier, $\Delta$ is so much smaller than any other dimensionless scale in the problem and in particular much smaller than the 
errors intrinsic in the effective theory and so can safely be ignored. Given that initially the measuring device has a unique ontic state $\ket{\psi(0)}$, the corresponding probability to be in the ontic state $\ket{\psi_i(T)}$ after the measurement is
\EQ{
p_i(T)=|c_i|^2+{\cal O}(\Delta)\ .
}
So the new interpretation  yields a satisfactory phenomenology in this simplest measurement model and, furthermore, yields the Born rule for the probabilities. There are, however, problems with the model:

(i) When the initial microscopic system has a degeneracy $|c_i|=|c_j|$ for some $i\neq j$. In the case of exact degeneracy one can easily show that the associated eigenvectors of $\hat\rho_\BA(T)$ after the measurement are no longer $\ket{\psi_i(T)}$ and $\ket{\psi_j(T)}$, but rather involves the macroscopic superpositions \cite{1993PhRvD..48.3768A}
\EQ{
\frac1{\sqrt2}\big(\ket{\psi_i(T)}\pm\ket{\psi_j(T)}\big)\ .
}
There is a superficial answer to this problem which says that a degeneracy of the state of $\BP$ requires infinite fine tuning that is unrealistic in practice \cite{JB}. Whilst this is true, one can show that even when there is no exact degeneracy the onset of decoherence can be delayed in an artificial way by having nearly degenerate states. If $|c_1|^2=\frac12+e^{-s}$ and $|c_2|^2=\frac12-e^{-s}$ then the decoherence time is proportional to $\sqrt{s}$ and so can be increased by tuning close to the degenerate point.

(ii) The model does not satisfy our requirement that the ontic states of macroscopic systems are robust against microscopic 
re-definitions of the sub-system. In this case, if we re-define $\BA$ to include the microscopic system $\BP$ itself then the ontic state of $A+P$ now becomes the pure state of the total system \eqref{b23} with probability 1. So the slightly re-defined system, with the addition of one microscopic degree-of-freedom, has led to a completely different ontology involving unacceptable superpositions of macroscopically distinct states. 

(iii) As it stands the model describes the case of a perfect measuring device that makes no errors. In realistic situations one could imagine that $\BA$ makes errors which manifests by less than perfect correlation between the states of $\BP$ and $\BA$ in \eqref{b23}. A more realistic situation has
\EQ{
\ket{\Psi(t)}=\sum_if_{ij}(t)\ket{\xi_i}\otimes\ket{\psi_j(t)}\ .
\label{b24}
}
with $f_{ij}(t)\to f^{(0)}_{ij}$, a constant, after the measurement time $t=T$.
Now the reduced density matrix of the measuring device is
\EQ{
\hat\rho_\BA(T)=\sum_{ijk}f_{ij}^{(0)}f^{(0)*}_{ik}\ket{\psi_j(T)}\bra{\psi_k(T)}
\label{cl2}
}
and since the matrix $f_{ij}^{(0)}$ need not be diagonal the ontic states will involve superpositions of the macroscopically distinct states $\ket{\psi_i(T)}$ \cite{albert1990wanted}.

\subsection{A More Realistic Model}\label{s2.2}

We now turn to a more sophisticated but realistic model which solves all three problems above. The new feature is that  the measuring device is also interacting with an environment $E$ and so the complete system is $\BP+\BA+E$. The initial state of the system is
\EQ{
\ket{\Psi(0)}=\Big[\sum_ic_i\ket{\xi_i}\Big]\otimes\ket{\Phi_0(0)}\ .
}
Here, $\ket{\Phi_0}$ is the initial state of the measuring device plus the environment
from which one deduces the reduced density matrix of $A$ by tracing out the environment:
\EQ{
\hat\rho_A(0)=\Tr_E\ket{\Phi_0(0)}\bra{\Phi_0(0)}\ ,
}
with eigenstates and eigenvalues
\EQ{
\hat\rho_A(0)\ket{\psi_a(0)}=p_a(0)\ket{\psi_a(0)}\ .
\label{yut}
}
We assume that the measuring device and environment are in equilibrium and so in a 
highly entangled state. In that case, the emergent classical view is ignorant of the exact ontic state of $A$. 
At equilibrium, the stochastic process described in section \ref{s1.3} is ergodic and it is
the macroscopic time average of this process that corresponds to what we think of as the initial classical state of the measuring device and this is captured by performing the ensemble average over $\hat\rho_A(0)$.

Evolving forward in time, and including measuring inefficiencies, we expect to have a state 
of the form
\EQ{
\ket{\Psi(t)}=\sum_{ij}f_{ij}(t)\ket{\xi_i}\otimes\ket{\Phi_j(t)}\ .
}
After the decoherence time $t>\tau$ and so certainly at the end of the measurement $t=T$, the states of $A+B$ are macroscopically distinct:
\EQ{
\big|\bra{\Phi_{i}(T)}\Phi_{j}(T)\rangle\big|\thicksim \Delta\ ,\qquad\text{for }i\neq j
\label{k23}
}
and $f_{ij}(t)\to f^{(0)}_{ij}$, a constant matrix.
Note that unitarity requires that the matrix $f_{ij}^{(0)}$ satisfies
\EQ{
\sum_{ij}\big|f_{ij}^{(0)}\big|^2=1+{\cal O}(\Delta)\ .
}
The reduced density matrix of the measuring device plus environment $A+E$ takes the form
\EQ{
\hat\rho_{A+E}(T)=\sum_{ijk}f_{ij}^{(0)}f_{ik}^{(0)*}\ket{\Phi_j}\bra{\Phi_k}\ ,
}
and so the ontic states of $A+E$ are 
\EQ{
\sum_jZ_{j}^{(i)}\ket{\Phi_j(T)}+{\cal O}(\Delta)\ ,
}
where the $Z_j^{(i)}$ are eigenvectors of the matrix with elements:
\EQ{
&M_{jk}=\sum_if_{ij}^{(0)}f^{(0)*}_{ik}\ ,\\ &\sum_jM_{jk}Z_k^{(i)}=\lambda^{(i)}Z_j^{(i)}\ .
}
So, unfortunately, the ontic states involve macroscopic superpositions of states of $A+E$ of the kind that we saw in the simple measurement model in the last section. 

However, if we examine the ontic states of $A$, which is the physically relevant sub-system, a more satisfactory picture emerges. Since the states of the environment are to a high degree orthogonal for $i\neq j$,
the reduced density matrix for $A$ takes the form
\EQ{
\hat\rho_A(T)=\sum_{ij}\big|f_{ij}^{(0)}\big|^2\hat\rho_A^{(j)}(T)+{\cal O}(\Delta)\ ,
\label{btt}
}
where we have defined the component reduced density matrices, one for each of the outcomes,
\EQ{
\hat\rho_A^{(j)}(T)=\Tr_E\ket{\Phi_j(T)}\bra{\Phi_j(T)}\ .
}
Since the states $\ket{\Phi_i(T)}$ have such a small inner product \eqref{k23}, the component density matrices commute to high degree of accuracy
\EQ{
[\hat\rho_A^{(i)}(T),\hat\rho_A^{(j)}(T)]={\cal O}(\Delta)\ .
}
The ontic states $\ket{\psi_a(T)}$, the eigenvectors of $\hat\rho_A(T)$, will, therefore, split up into $d_P$ mutually ergodically inaccessible sets
${\cal E}^{(i)}$, as well as a set ${\cal E}_0$ of approximately null eigenvectors that play no important role. The component density matrices take the form
\EQ{
\hat\rho^{(i)}_A(T)=\sum_{a\in{\cal E}^{(i)}}p_a^{(i)}\ket{\psi_a(T)}\bra{\psi_a(T)}+{\cal O}(\Delta)\ .
}
Therefore the ontic state $\ket{\psi_a(T)}$, $a\in{\cal E}^{(j)}$, has $\hat\rho_A(T)$ eigenvalue
\EQ{
p_a(T)=\sum_{i}\big|f_{ij}^{(0)}\big|^2p_a^{(j)}+{\cal O}(\Delta)\ .
\label{nw2}
}
It does not involve a macroscopic superposition---at least to order $\Delta$---and so
compared to \eqref{cl2}, one can see that the environment has the effect of yielding an acceptable phenomenology.

Under the stochastic process, even for macroscopic time scales, 
there is only a minute probability for $\ket{\psi_{a}(T)}$ to make a transition into another state $\ket{\psi_{b}(T)}$ in a different ergodic set $a\in {\cal E}^{(i)}$ and $b\in{\cal E}^{(j)}$,
for $i\neq j$. For fixed $j$, the group of ontic states $\ket{\psi_{a}(T)}$, $a\in{\cal E}^{(j)}$, are not macroscopically distinct, and so ontic states can effectively only make transitions within this set. We can, in principle, use the stochastic process to work out the probability $p_{a|b}(T,0)$ that an initial ontic state $\ket{\psi_b(0)}$ evolves to $\ket{\psi_{a}(T)}$ using \eqref{s99}. However, according to our discussion below equation \eqref{yut}, since the measuring device is assumed to be in equilibrium with the environment its ontic states are undergoing constant transitions. The time average of this process is captured by taking the ensemble average over $\hat\rho_A(0)$.
With this initial averaging, we can use \eqref{s99} to compute the probabilities that the system is finally in the ontic state $\ket{\psi_a(T)}$:
\EQ{
\sum_bp_{a|b}(T,0)p_b(0)=p_a(T)\ .
}
The result is simply the single-time probability in \eqref{nw2}. Turning to the final state,
as far as the emergent classical ontology is concerned, the ontic states in an ergodic set ${\cal E}^{(j)}$ cannot be distinguished and are also undergoing constant transitions between themselves. Therefore it is only meaningful to compute the inclusive probability that the system ends up in the $j^\text{th}$ set:
\EQ{
p^{(j)}(T)=\sum_{a\in{\cal E}^{(j)}}p_a(T)=\sum_i\big|f_{ij}^{(0)}\big|^2+{\cal O}(\Delta)\ ,
\label{bfr}
}
where we used the fact that $\Tr\hat\rho_A^{(j)}=1$ so that $\sum_{a\in{\cal E}^{(j)}}p_a^{(j)}=1+{\cal O}(\Delta)$. It is particularly noteworthy that the final result here is actually independent of the detailed form of the microscopic stochastic process that we introduced in section \ref{s1.3} because we averaged over the initial ontic state and the calculated an inclusive probability for a sub-ensemble in the final state. The non-trivial role of the stochastic process is that it was defined in such a way so as to lead to ergodicity breaking. 
Note that for a perfect measuring device $f_{ij}(t)=c_i\delta_{ij}$ and so $p^{(j)}(T)=|c_j|^2$ which is just the Born rule.

Notice that, although the individual probabilities $p_a(t)$ can depend on time as a result of the interaction of $\BA$ with the environment, the sum over the group is independent of time to order $\Delta$, assuming, as we have, that the measuring device settles down and ceases to make errors for $t>T$. 
It is clear that the more realistic model solves problems (i) and (iii). 
Firstly, (iii) is solved because the ontic states $\ket{\psi_a(T)}$ do not involve macroscopic superpositions. In addition, (i) is solved because the near degeneracy $|c_i|\approx|c_j|$ no longer leads to macroscopic superpositions or artificially extended de-coherence times. 

Now we turn to problem (ii). The reduced density matrix of $\BA+\BP$ is
\EQ{
\hat\rho_{\BA+\BP}(T)=\sum_{ij}\big|f_{ij}^{(0)}\big|^2\ket{\Xi^{(j)}}\bra{\Xi^{(j)}}\otimes\hat\rho_A^{(j)}(T)
+{\cal O}(\Delta)\ ,
}
where
\EQ{
\ket{\Xi^{(j)}}=\Big[\sum_i\big|f_{ij}^{(0)}\big|^2\Big]^{-1/2}\sum_if_{ij}^{(0)}\ket{\xi_i}
}
and so the ontic states are
\EQ{
\ket{\Xi^{(j)}}\otimes\ket{\psi_a(T)}+{\cal O}(\Delta)\ ,\quad a\in{\cal E}^{(j)}\ .
\label{gg5}
}
The probabilities are identical to \eqref{nw2}. Note that the ontic states \eqref{gg5} are those of $\BA$ not entangled with states of $\BP$. Hence, focussing on $\BA+\BP$ rather than $\BA$ does not lead to the disastrous change in the ontology seen in the simple model.
However, unless the measuring device is perfect, the states $\ket{\Xi^{(j)}}$ are not the ontic states of $P$, they are not even orthogonal. 

\subsection{Collapse of the Wave Function}\label{s2.3}

After the measurement has been completed, the stochastic process ensures that  there are effectively no transitions between the states in different ergodic sets.
As far as $\BA$ is concerned, if the state $\ket{\psi_a(T)}$ has $a\in{\cal E}^{(j)}$ then for all practical purposes for calculating future dynamics, it would be prudent, but not necessary, to remove the other terms $i\neq j$ from $A$'s reduced density matrix 
\eqref{btt} leading to the replacement
\EQ{
\hat\rho_\BA(T)~\rightsquigarrow~
\hat\rho_\BA^{(j)}(T)\ .
\label{spp2}
}
So the collapse of the wave function in the new interpretation  is just the innocuous process of removing
terms which could only have an effect of order $\Delta$ in the future. As we have repeatedly emphasised such effects are many orders of magnitude smaller than other kinds of systematic errors that are inherent in the model. So if we condition the future of a system on its present state---the ergodic sub-ensemble of ontic states at a given time---then collapse of the wave function is a harmless procedure of removing terms that are ergodically inaccessible. 

\subsection{Measurement of Continuous Quantum Systems}\label{s2.4}

Measurements on microscopic systems with continuous eigenvectors, like the position of a particle, have caused problems for modal interpretations \cite{BacciagaluppiDonaldVermaas:1995cddpmi,Donald:1998dcdpmi,Page:2011gi}. In this case one might expect a generalization of \eqref{b23} to be
\EQ{
\ket{\Psi(t)}=\int dx\,\xi(x)\ket{x}\otimes\ket{\psi_x(t)}\ .
\label{b25}
}
Here, $\ket{\psi_x(t)}$ is a macroscopic state of $\BA$ indicating that the particle is at $x$.
The reduced density matrix of the measuring device is then
\EQ{
\hat\rho_\BA(t)=\int dx\,|\xi(x)|^2\ket{\psi_x(t)}\bra{\psi_x(t)}\ .
}
The potential problem is that even though the states $\ket{\psi_x(t)}$ are expected to become
approximately orthogonal, for instance, one expects
\EQ{
\big|\bra{\psi_x(T)}\psi_y(T)\rangle\big|\thicksim\exp\big[-\lambda N(x-y)^2\big]\ ,
}
the eigenvectors of $\hat\rho_\BA(T)$, contrary to na\"\i ve expectations, are not localized in $x$. In fact, if the original wave function of the particle $\xi(x)$ is spread out over a range $\delta x$ then the ontic states of the measuring device 
involve a spread of the states $\ket{\psi_x(t)}$ over the same range.
In other words, the ontic states of $\BA$ would involve macroscopic superpositions. This is obviously a potential disaster for any modal interpretation.

The loop-hole in this thinking was identified in \cite{Hollowood:2013cbr}. In fact
we should be on alert for any argument, like that above, that relies on the smoothness of wave functions down to be arbitrarily small scales as this is not realistic and in the spirit of 
effective theory. In fact non-relativistic quantum mechanics based on the one-particle truncation of the multi-particle Hilbert space breaks down on length scales $\hbar/mc$. So any attempt to measure the position of the particle down to these scale will inevitably involve particle creation and annihilation. But in a more practical sense, any
realistic description would acknowledge the fact that $A$ would have some inherent finite resolution scale. A simple way to model this is to imagine that the states of $A$ respond to the particle's position in finite bins $[x_j,x_{j+1}]$, where $x_j=x_0+j\epsilon$, where $\epsilon$ is the resolution scale. As along as $\epsilon\gg\hbar/mc$ the description of the measurement process within non-relativistic quantum mechanics will be valid.
In that case \eqref{b25} should be replaced by 
\EQ{
\ket{\Psi(t)}=\sum_j\int_{x_j}^{x_{j+1}} dx\,\xi(x)\ket{x}\otimes\ket{\psi_j(t)}\ .
\label{b26}
}
After the measurement, the ontic states of $\BA$ are approximately one of the discrete macroscopically distinct states $\ket{\psi_j(t)}$ and so a satisfactory phenomenology without macroscopic superpositions ensues. 

In \cite{Hollowood:2013cbr} a similar issue was shown to arise in an experiment involving the monitoring of a decaying system. In this case if the measuring device is taken to have infinite temporal resolution then superpositions of macroscopically distinct states arise. However, once the finite temporal resolution scale of a realistic measuring device is taken into account the problem with macroscopic superpositions evaporates. 

\section{EPR-Bohm and Bell}\label{s4.1}

In this section, we discuss the new interpretation  in the context of Bohm's classic 
thought experiment \cite{Bohm:1951qt,BohmAharonov:1957depperp}, based originally
on the classic paper by Einstein, Podolsky, and Rosen \cite{EinsteinPodolskyRosen:1935cqmdprbcc}, and the implications for Bell's theorem \cite{Bell:1964oeprp}. 

It is useful for simplicity to avoid introducing a separate environment. 
Our experience from section \ref{s2.1} suggests that we may do this as long as (i) we avoid degeneracies in the state of microscopic system (ii) have perfect measuring devices. The point is that the micro-system acts as a surrogate environment for the measuring devices and the ontic states of the measuring devices act as proxies for the ergodic sets of ontic states of the more realistic situation with a large environment. 

\subsection{The Thought Experiment}

The EPR-Bohm set-up begins with a pair of qubits initially prepared in the entangled state that we take to be
\EQ{
\ket{\Phi}=c_+\ket{z^+z^-}+c_-\ket{z^-z^+}\ .
}
Note that we assume that $c_\pm$ are generic to avoid degeneracies.
The ontic states of 1 or 2 are each one of the pair $\ket{z^\pm}$ with probabilities $|c_\pm|^2$, or vice-versa, respectively. 
\begin{figure}[ht]
\begin{center}
\begin{tikzpicture}[scale=0.65]
\node at (4.3,0) (a1) {$B$};
\node at (-3.3,0) (a1) {$A$};
\node at (1.5,0.4) (a1) {$2$};
\node at (-1,0.4) (a1) {$1$};
\filldraw[black] (0,0) circle (0.5cm);
\draw[->,very thick] (0,0) -- (3,0);
\draw[->,very thick] (0,0) -- (-2,0);
\begin{scope}[xshift=3cm]
\draw[very thick] (0.2,0.3) -- (0.7,0.3) -- (0.7,-0.3) -- (0.2,-0.3);
\draw[->] (0,-0.9) -- (1,0.9);
\node at (1.3,1.3) (b1) {$m$};
\end{scope}
\begin{scope}[xshift=-2cm,rotate=180]
\draw[very thick] (0.2,0.3) -- (0.7,0.3) -- (0.7,-0.3) -- (0.2,-0.3);
\draw[<-] (0.8,-0.9) -- (0.2,0.9);
\node at (0.9,-1.3) (b1) {$n$};
\end{scope}
\end{tikzpicture}
\end{center}
\caption{\small The EPR-Bohm thought experiment. Two qubits in the entangled state $\ket{\Phi}$ are produced at the source and then recoil back-to-back towards 2 qubit detectors $A$ and $B$ designed to measure the component of the spin along directions $n$ and $m$, respectively. In our set up, we choose an inertial frame for which the interaction between $A$ and 1 happens before $B$ and 2.}
\label{f18}
\end{figure}
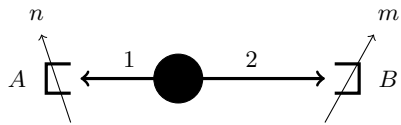

The ontic state of $1+2$ is uniquely $\ket{\Phi}$. Given our discussion in section \ref{s1.4}, we can say that the systems $1$, $2$ and $1+2$ are related in a ``quantum" way and there is no notion of a joint property assignment or associated probability for the ontic states of 1 and 2. 
 
In order to perform measurements, we
add two spatially-separated spin detectors $A$ and
$B$, where $A$ detects the spin of particle $1$ and $B$ detects
the spin of particle $2$ as in figure \ref{f18}.
The complete system consists of qubits 1 and 2, measuring devices $A$ and $B$ and the environment, although as mentioned above in the present context we will ignore the environment in this simplified analysis. It is important that even though there is no genuine environment, the ontic state of $A+B$ is always related to $A$ and $B$ in a classical way, that is as a tensor product of the ontic states of $A$ and $B$ as in \eqref{mdr} but in this simplified model exactly. So the description from the point-of-view of $A$, $B$ and $A+B$ can be integrated into a consistent global description throughout the experiment. Another important point is that the mirror ontic states to those of $A+B$ are the ontic states of $1+2$ themselves so that the ontic dynamics of $A+B$ depends implicitly on $1+2$ as is clear from the matrix elements in \eqref{frr}.

Let us analyse what happens if $A$ is set to measure the spin component in the $n$ direction, at an angle $\theta$ to the $z$-axis in the $(x,z)$ plane, and $B$ is set to measure the spin in a direction $m$ at an angle $\phi$ to the $z$-axis in the $(x,z)$ plane.\footnote{The eigenstates are $\ket{n^\pm}=\cos(\theta/2)\ket{z^\pm}\pm\sin(\theta/2)\ket{z^\mp}$ and  $\ket{m^\pm}=\cos(\phi/2)\ket{z^\pm}\pm\sin(\phi/2)\ket{z^\mp}$.} This set-up is generic enough for our purposes. 

The initial state of the overall system is
\EQ{
\ket{\Psi(t_1)}=\ket{A_0 B_0}\otimes\ket{\Phi}\ .
\label{sw1}
}
Suppose, in a certain inertial frame, the interaction between $A$ and 1 happens first at time $t_A>t_1$. After a short decoherence time, the state of $1$ becomes entangled with $A$ and the state of the total system becomes for $t_2>t_A$
\EQ{
\ket{\Psi(t_2)}=\ket{A_+B_0}\otimes\ket{n^+\psi^+}+\ket{A_-B_0}\otimes\ket{n^-\psi^-}\ ,
}
where
\EQ{
\ket{\psi^\pm}=c_\pm\cos(\theta/2)\ket{z^\mp}\pm c_\mp\sin(\theta/2)\ket{z^\pm}\ .
\label{l23}
}
Note that these states are neither normalised or orthogonal.

Assuming in this simple model that the states $\ket{A_\pm}$ are exactly orthogonal, the ontic state of $1+2$ has changed from $\ket{\Phi}$ to either of the non-entangled (and non-normalized states) $\ket{n^\pm\psi^\pm}$
with probabilities $\bra{\psi^\pm}\psi^{\pm}\rangle$. The ontic state of $1$ is now one of the pair $\ket{n^\pm}$ and these are now perfectly matched with the ontic states of $1+2$.

The ontic states of $1+2$, after the local interaction between $A$ and 1, are, of course, the analogues of the collapsed wave functions of the Copenhagen interpretation. So the puzzle seems to be that a local interaction between $A$ and $1$ has changed the ontic state of $1+2$ and this seems to have led to a non-local change of the ontic state of qubit 2.
But it turns out this is not correct.
The key point point is that the interaction between $A$ and 1 changes the ontic states of 1 and $1+2$ but not 2.
To see this, the reduced density matrix of $2$ before the interaction between $A$ and $1$ is
\EQ{
\hat\rho_2(t_1)=|c_+|^2\ket{z^-}\bra{z^-}+|c_-|^2\ket{z^+}\bra{z^+}\ ,
}
while after the interaction,
\EQ{
\hat\rho_2(t_2)=\ket{\psi^+}\bra{\psi^+}+\ket{\psi^-}\bra{\psi^-}\ .
}
But the states $\ket{\psi^\pm}$ are neither normalized or orthogonal
and using \eqref{l23} one can show that actually
\EQ{
\hat\rho_2(t_1)\equiv\hat\rho_2(t_2)\ ,
}
reflecting the fact that ontic assignments are local and an interaction between $A$ and 1 cannot affect the ontic state of the causally separated qubit 2. The implication is that the 
ontic state of qubit 2 remains one of the pair $\ket{z^\pm}$ throughout the interaction between $A$ and qubit 1. To be clear, the interaction between $A$ and $1$ does not change the possible ontic states $\ket{z^\pm}$ of 2 but also the actual ontic state of 2. The latter fact follows from the stochastic process describing the ontic states of $2$ given that the interaction Hamiltonian $\hat H_\text{int}$ is itself local, that is of the form of a sum of two local interactions: 
\EQ{
\hat H_\text{int}=\hat H^{A,1}_\text{int}+\hat H^{B,2}_\text{int}\ ,
\label{ihl}
}
where the first (second) term is non-vanishing for $t$ in the neighbourhood of $t_A$ ($t_B$). This implies
\EQ{
p_{z^\pm|z^\pm}(t_2,t_1)=1\ .
}
So although the interaction between $A$ and $1$ changes the ontic state of $1+2$ from $\ket{\Phi}$ to one of the pair $\ket{n^\pm\psi^\pm}$ and this seems to have fundamentally changed the state of $2$, however, the ontic states of $1+2$ cannot be broken down in terms of the ontic states of its sub-systems $1$ and $2$---at least for $t<t_B$. Consequently, there is an apparent duality for qubit 2 from the point-of-view of $1+2$ compared with 2:
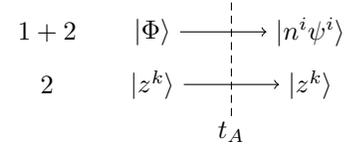
\EQ{
\begin{tikzpicture}[xscale=0.7,yscale=0.7]
\node at (-2,0) (z1) {$2$};
\node at (-2,1) (z2) {$1+2$};
\node at (0,1) (b1) {$\ket{\Phi}$};
\node at (3,1) (b2) {$\ket{n^i\psi^{i}}$};
\node at (0,0) (c1) {$\ket{z^k}$};
\node at (3,0) (c2) {$\ket{z^k}$};
\draw[->] (b1) -- (b2);
\draw[->] (c1) -- (c2);
\draw[-,densely dashed] (1.5,-0.6) -- (1.5,1.6);
\node at (1.5,-0.9) (z1) {$t_A$};
\end{tikzpicture}\notag
}
It is tempting to say that the behaviour of the state of $1+2$ here is non-local. However, this is 
potentially misleading because in order to talk about locality within the system $1+2$ requires us to break $1+2$ down in terms of its sub-systems 1 and 2. But, to reiterate, we cannot break down the states $\ket{n^i\psi^{i}}$ ontically in terms of $1$ and $2$ because the descriptions via $1+2$ and $2$ cannot be integrated into a consistent whole. We will have more to say about this and its implications in section \ref{10.2}.

\begin{figure}[ht]
\begin{center}
\begin{tikzpicture}[xscale=1,yscale=1]
\begin{scope}[yshift=-2cm]
\node at (1,10) (b1) {$1$};
\node at (2,10) (b2) {$\ket{z^\pm}$};
\node at (4,10.2) (b3) {$\ket{n^+}$};
\node at (4,9.8) (b4) {$\ket{n^-}$};
\node at (6,10.2) (b5) {$\ket{n^+}$};
\node at (6,9.8) (b6) {$\ket{n^-}$};
\draw[-] (b2) -- (3,10);
\draw[->] (3,10) -- (b3);
\draw[->] (3,10) -- (b4);
\draw[->] (b3) -- (b5);
\draw[->] (b4) -- (b6);
\end{scope}
\begin{scope}[yshift=-3cm]
\node at (1,10) (b1) {$2$};
\node at (2,10) (b2) {$\ket{z^\pm}$};
\node at (4,10) (b3) {$\ket{z^\pm}$};
\node at (6,10.2) (b5) {$\ket{m^+}$};
\node at (6,9.8) (b6) {$\ket{m^-}$};
\draw[->] (b2) -- (b3);
\draw[-] (b3) -- (4.9,10);
\draw[->] (4.9,10) -- (b5);
\draw[->] (4.9,10) -- (b6);
\end{scope}
\begin{scope}[yshift=-4.4cm]
\node at (1,10) (b1) {$1+2$};
\node at (2,10) (b2) {$\ket{\Phi}$};
\node at (4,10.2) (b3) {$\ket{n^+\psi^+}$};
\node at (4,9.8) (b4) {$\ket{n^-\psi^-}$};
\node at (6,10.6) (b5) {$\ket{n^+m^+}$};
\node at (6,10.2) (b6) {$\ket{n^+m^-}$};
\node at (6,9.8) (b7) {$\ket{n^-m^+}$};
\node at (6,9.4) (b8) {$\ket{n^-m^-}$};
\draw[-] (b2) -- (3,10);
\draw[->] (3,10) -- (b3);
\draw[->] (3,10) -- (b4);
\draw[-] (b3) -- (4.9,10.2);
\draw[->] (4.9,10.2) -- (b5);
\draw[->] (4.9,10.2) -- (b6);
\draw[-] (b4) -- (4.9,9.8);
\draw[->] (4.9,9.8) -- (b7);
\draw[->] (4.9,9.8) -- (b8);
\end{scope}
\draw[-,densely dashed] (3,8.5) -- (3,4.5);
\draw[-,densely dashed] (4.9,8.5) -- (4.9,4.5);
\node at (3,4) (z1) {$t_A$};
\node at (4.9,4) (z1) {$t_B$};
\node at (2,4.5) (z1) {quantum};
\node at (3.9,4.5) (z1) {quantum};
\node at (5.8,4.5) (z1) {classical};
\end{tikzpicture}
\end{center}
\caption{\small Snapshots of the ontic states of 1, 2 and $1+2$ at times $t_1<t_A<t_2<t_B<t_3$ in the given inertial frame. For $t=t_1$ and $t_2$, the ontic states of $1+2$ are not tensor products of the ontic states of 1 and 2 and so the ontology is ``quantum". In the last time step $t=t_3$ the relation between states becomes ``classical" in the sense that the ontic states of $1+2$ are tensor products of those of 1 and 2.}
\label{f3}
\end{figure}
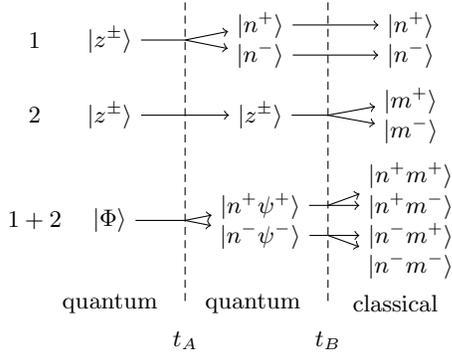

To emphasize, for $t<t_B$, the ontic state of $1+2$ is not a tensor product of ontic states of 1 and 2 so there is no sense in which the ontology is ``classical" in the sense described in section \ref{s1.4}. So in figure \ref{f3}, which summarizes the ontic dynamics of 1, 2 and $1+2$, we label it as ``quantum".

\begin{figure}[ht]
\begin{center}
\begin{tikzpicture}[xscale=1,yscale=1]
\begin{scope}[yshift=-2cm]
\node at (0.6,10) (b1) {$A$};
\node at (2,10) (b2) {$\ket{A_0}$};
\node at (4,10.2) (b3) {$\ket{A_+}$};
\node at (4,9.8) (b4) {$\ket{A_-}$};
\node at (6,10.2) (b5) {$\ket{A_+}$};
\node at (6,9.8) (b6) {$\ket{A_-}$};
\draw[-] (b2) -- (3,10);
\draw[->] (3,10) -- (b3);
\draw[->] (3,10) -- (b4);
\draw[->] (b3) -- (b5);
\draw[->] (b4) -- (b6);
\end{scope}
\begin{scope}[yshift=-3cm]
\node at (0.6,10) (b1) {$B$};
\node at (2,10) (b2) {$\ket{B_0}$};
\node at (4,10) (b3) {$\ket{B_0}$};
\node at (6,10.2) (b5) {$\ket{B_+}$};
\node at (6,9.8) (b6) {$\ket{B_-}$};
\draw[->] (b2) -- (b3);
\draw[-] (b3) -- (4.9,10);
\draw[->] (4.9,10) -- (b5);
\draw[->] (4.9,10) -- (b6);
\end{scope}
\begin{scope}[yshift=-4.4cm]
\node at (0.7,10) (b1) {$A+B$};
\node at (2,10) (b2) {$\ket{A_0B_0}$};
\node at (4,10.2) (b3) {$\ket{A_+B_0}$};
\node at (4,9.8) (b4) {$\ket{A_-B_0}$};
\node at (6,10.6) (b5) {$\ket{A_+B_+}$};
\node at (6,10.2) (b6) {$\ket{A_+B_-}$};
\node at (6,9.8) (b7) {$\ket{A_-B_+}$};
\node at (6,9.4) (b8) {$\ket{A_-B_-}$};
\draw[-] (b2) -- (3,10);
\draw[->] (3,10) -- (b3);
\draw[->] (3,10) -- (b4);
\draw[-] (b3) -- (4.9,10.2);
\draw[->] (4.9,10.2) -- (b5);
\draw[->] (4.9,10.2) -- (b6);
\draw[-] (b4) -- (4.9,9.8);
\draw[->] (4.9,9.8) -- (b7);
\draw[->] (4.9,9.8) -- (b8);
\end{scope}
\draw[-,densely dashed] (3,8.5) -- (3,4.5);
\draw[-,densely dashed] (4.9,8.5) -- (4.9,4.5);
\node at (3,4) (z1) {$t_A$};
\node at (4.9,4) (z1) {$t_B$};
\node at (2,4.5) (z1) {classical};
\node at (3.9,4.5) (z1) {classical};
\node at (5.8,4.5) (z1) {classical};
\end{tikzpicture}
\end{center}
\caption{\small Snapshots of the ontic states of the measuring devices $A$, $B$ and $A+B$ at $t_1$, $t_2$ and $t_3$. The relation between the three systems $A$, $B$ and $A+B$ is always classical, as the ontic states of $A+B$ are just tensor products of the ontic states of $A$ with those of $B$.
However, for $t=t_3$, the measuring devices are, nevertheless, correlated in a classical sense because $\hat\rho_{A+B}\neq\hat\rho_A\otimes\hat\rho_B$.}
\label{f4}
\end{figure}
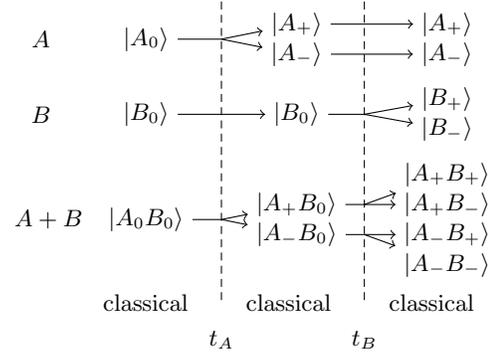

On the other hand the behaviour of $A+B$, $A$ and $B$ is always classical as illustrated in figure \ref{f4}. The initial ontic state of $A+B$ is $\ket{A_0B_0}$ while after the measurement at $A$ the ontic state is then one of the pair $\ket{A_\pm B_0}$ with probabilities
\EQ{
p_{A_\pm B_0|A_0B_0}(t_2,t_1)=\bra{\psi^\pm}\psi^\pm\rangle\ ,
\label{m23}
}
for $t_1<t_A<t_2$. These follow from applying \eqref{pc6} along with the uniqueness of the initial ontic state.

Finally, at time $t_B$ in this frame, $B$ interacts with qubit 2 and after a further short decoherence time, the state becomes
\EQ{
\ket{\Psi(t_3)}=\sum_{ij=\pm}\ket{A_iB_j}\otimes\ket{n^im^j}\bra{m^j}\psi^{i}\rangle\ .
}
After this interaction the 
triplet of systems 1, 2 and $1+2$ now has ``classical" ontology in the sense that the ontic states of $1+2$ are one of the quartet $\ket{n^im^j}$ and these are tensor products of the ontic states of $1$ and $2$. For the measuring devices, the final ontic states are one of the quartet $\ket{A_iB_j}$. It is important is that (i) the first interaction between $A$ and $1$ does not change the ontic state of $B$ and (ii) the second interaction between $B$ and $2$ does not change the ontic state of $A$. This is guaranteed by the locality of the interaction Hamiltonian \eqref{ihl}:
\EQ{
p_{B_0|B_0}(t_2,t_1)=1\ ,\qquad p_{A_i|A_j}(t_3,t_2)=\delta_{ij}\ .
}
The implication is that the ontic dynamics has the tree-like structure as shown in figure \ref{f4}. Therefore, once again using \eqref{pc6}, the only non-vanishing probabilities are
\EQ{
p_{A_iB_j|A_iB_0}(t_3,t_2)=\frac{\big|\bra{m^j}\psi^{i}\rangle\big|^2}{\bra{\psi^{i}}\psi^{i}\rangle}\ ,
\label{m24}
}
for $t_2<t_B<t_3$.
The final probabilities follow from \eqref{pc6}, or by composing \eqref{m23} and \eqref{m24},
\EQ{
p_{A_iB_j|A_0B_0}(t_3,t_1)=\big|\bra{m^j}\psi^{i}\rangle\big|^2\ .
}
As we have emphasised,
the ontic states of $A$, $B$ and $A+B$ are just tensor products of those of $A$ and $B$ for all $t$.
In the more realistic model with an environment, this relation will be emergent as in \eqref{mdr}.
Given the classical ontology, it is meaningful to define joint probabilities as in \eqref{jpb}:
\EQ{
p(A_i B_j)\overset{\text{emergent}}=p_{A_i B_j|A_0B_0}(t_3,t_1)\ .
\label{jqq}
}
These emergent probabilities are exactly what would have been predicted on the basis of Born's rule. So during the experiment the entanglement between 1 and 2 is converted into an emergent classical correlation between $A$ and $B$. But what is crucial is that this correlation arises after the {\it local\/} interaction between $1$ and $A$ and separately between $2$ and $B$ and not through any non-local interaction between $A$ and $B$.

\subsection{Bell's Theorem}\label{10.2}

Now we turn to a discussion of Bell's theorem \cite{Bell:1964oeprp,Bell:2004suqm}. The first point is that, since the predictions agree with the conventional analysis of the Copenhagen interpretation and the Born rule, it must be that the new interpretation implies a violation of the Bell inequality. 

In fact it is easy to see that the new interpretation violates {\it outcome independence\/}:
\EQ{
p(B_j|A_i,n,m)=p(B_j|n,m)\ .
\label{u23}
}
To see this note that the conditional probability above can only be defined in the context of the ontic dynamics of sub-system $A+B$; indeed, 
\EQ{
p(B_j|A_i,n,m)&\equiv p_{A_i B_j|A_iB_0}(t_3,t_2)\\ &
=\frac{\big|\bra{m^j}\psi^{i}\rangle\big|^2}{\bra{\psi^{i}}\psi^{i}\rangle}\ .
\label{x89}
}
We can now pinpoint where the dependence on the state of $A$ arises in the interaction between $B$ and $2$. Although the sub-system $A+B$ can be broken down in terms of its component parts $A$ and $B$ because their ontology is aways ``classical", the ontic dynamics of $A+B$ depends on the mirror ontic state of $1+2$, and $1+2$ does not have a ``classical" ontology for $t<t_B$. The relation between the ontic states of $A+B$ and the associated mirror ontic states of  $1+2$ are as follows:
\EQ{
\begin{tikzpicture}[xscale=0.7,yscale=0.7]
\node at (0,1.6) (a1) {$\ket{A_0B_0}$};
\node at (3,1.6) (a2) {$\ket{A_i B_0}$};
\node at (6,1.6) (a3) {$\ket{A_i B_j}$};
\node at (0,0) (b1) {$\ket{\Phi}$};
\node at (3,0) (b2) {$\ket{n^i \psi^{i}}$};
\node at (6,0) (b3) {$\ket{n^i m^j}$};
\draw[->] (a1) -- (a2);
\draw[->] (a2) -- (a3);
\draw[->] (b1) -- (b2);
\draw[->] (b2) -- (b3);
\node[rotate=90] at (0,0.8) (c1) {$\leftrightsquigarrow$};
\node[rotate=90] at (3,0.8) (c1) {$\leftrightsquigarrow$};
\node[rotate=90] at (6,0.8) (c1) {$\leftrightsquigarrow$};
\draw[-,densely dashed] (1.4,-0.6) -- (1.4,2);
\draw[-,densely dashed] (4.5,-0.6) -- (4.5,2);
\node at (1.4,-0.9) (z1) {$t_A$};
\node at (4.5,-0.9) (z1) {$t_B$};
\end{tikzpicture}\notag
}
It is the entanglement of the mirror ontic state $\ket{\Phi}$ that leads, after the first interaction at $t=t_A$, to the mirror ontic states $\ket{n^i\psi^{i}}$. Including only the important tensor product factors, we have
\EQ{
\begin{tikzpicture}[xscale=0.7,yscale=0.7]
\node at (-0.5,-1.6) (a3) {$\hat H^{A,1}_\text{int}$};
\node at (0,0) (b1) {$\ket{A_0}\otimes\ket{\Phi}$};
\node at (4.65,0) (b2) {$\ket{A_i}\otimes\ket{n^i}\otimes\ket{\psi^{i}}$};
\draw[->] (a3) -- (0.3,-0.5);
\draw[->] (a3) -- (-0.7,-0.5);
\draw[->] (b1) -- (b2);
\draw[-,densely dashed] (1.85,-0.6) -- (1.85,0.6);
\node at (1.85,-0.9) (z1) {$t_A$};
\end{tikzpicture}\notag
}
It is important to remember that $\ket{\psi^{i}}$ are not the ontic states of qubit 2 even though they are tensor product factors of the ontic states of $1+2$ for $t=t_2$.
This in turn implies that when $B$ interacts with qubit 2 at $t_B$, the ontic dynamics becomes implicitly dependent on the ontic state of $A$ through the mirror ontic state $\ket{n^i\psi^{i}}$. In this interaction, locality ensures that only the tensor product factors of $B$ and qubit 2 are relevant but the state of $A$ influences the dynamics via this initial tensor product factor of qubit 2, that is $\ket{\psi^{i}}$, and this leads to the $A_i$ dependent probabilities \eqref{x89}:
\EQ{
\begin{tikzpicture}[xscale=0.7,yscale=0.7]
\node at (0,-1.6) (a4) {$\hat H^{B,2}_\text{int}$};
\draw[->] (a4) -- (1.1,-0.5);
\draw[->] (a4) -- (-0.1,-0.5);
\node at (0,0) (b1) {$\ket{A_i}\otimes\ket{B_0}\otimes\ket{\psi^{i}}$};
\node at (5.5,0) (b2) {$\ket{A_i}\otimes\ket{B_j}\otimes\ket{m^j}$};
\draw[->] (b1) -- (b2);
\draw[-,densely dashed] (2.75,-0.6) -- (2.75,0.6);
\node at (2.75,-0.9) (z1) {$t_B$};
\end{tikzpicture}\notag
}
Note that during the interaction between $B$ and qubit 2 the state of causally separated $\ket{A_i}$ is inert but is included in the above to see the correlation with the states $\ket{\psi^{i}}$.

On the other hand, the probabilities satisfy {\it parameter independence\/}:
\EQ{
p(A_\pm|n,m)=p(A_\pm|n)\ ,
\label{k24}
}
since
\EQ{
p(A_\pm|m,n)&\equiv\sum_{j=\pm}p_{A_\pm B_j|A_0B_0}(t_3,t_1)\\ &=|c_\pm|^2
\cos^2(\theta/2)+|c_\mp|^2\sin^2(\theta/2)\ ,
}
independent of $m$, i.e.~$\phi$. This latter condition expresses the fact that
what happens at $A$ does not depend on what is measured at $B$.
The violation of outcome independence but observance of parameter independence is just as in the Copenhagen interpretation and a violation of either implies a violation of the Bell inequality. So what we have shown is that the new interpretation violates Bell's inequality for the same reason the Copenhagen interpretation does because the initial state of the qubits $\ket{\Phi}$ is entangled. The novelty is that it does this without invoking the collapse of the wave function. Instead what happens is a curious and apparently non-local change in the ontic state of $1+2$ from $\ket{\Phi}$ to $\ket{n^i\psi^i}$ when $A$ and $1$ interact. On closer inspection, though, it is hard to say this is a non-local process because it does not make sense to analyse the states $\ket{n^i\psi^i}$ in terms of the ontic states of their constituent qubits. So this is the ``spooky action at a distance" of quantum mechanics laid bare. In any event there is no violation of causality.

Another point to make is that the assignment of ontic states clearly depends on the inertial frame chosen to view the experiment. For instance in our chosen frame, where the measurement is made at $A$ before $B$, the ontic states of $A+B$ after the first measurement are $\ket{A_\pm B_0}$. But in another frame, where the measurement is made at $B$ before $A$, the ontic states of $A+B$ after the first measurement are $\ket{A_0B_\pm}$. This, of course, is no surprise, but modal interpretations have been claimed to be inconsistent with special relativity \cite{myr}. Fortunately these claims have shown to rest on the false assumption that the ontic states of a sub-system are global property assignments and so are unwarranted \cite{BH}.
 
\section{Discussion}\label{s5}

The emergent Copenhagen interpretation is a completely self-contained interpretation of the quantum mechanics that builds on and modifies earlier proposals that are collectively known as  modal interpretations. The key ingredient is the fact that one cannot isolate a quantum system from its environment. The interaction between the two allows one to define the dual notion of the epistemic and ontic state of the sub-system. The former evolves according to the Schr\"odinger equation (for the total system), whilst the latter evolves according to a stochastic process. This dualism is analogous to the ensemble and micro-state of classical statistical mechanics and allows for the solution of the measurement problem by invoking an ergodicity argument familiar from the discussion of a phase transition. Although we have presented the quantum interpretation as an analogue of classical statistical mechanics, in fact the former should be taken as a proper definition of the latter. In this regard, we are taking the viewpoint of \cite{PopescuShortWinter:2005fsmeisa,PopescuShortWinter:2006efsm,BL,LL,GoldsteinLebowitzTumulkaZanghi:2006ct,LPSW,Sh,GMM} but reintroducing the notion of a micro-state in the form of the ontic state. However, unlike the situation of micro-states and ergodicity in classical statistical mechanics, the dynamics of ontic states follows from a very simple Markov process.

The most striking feature of the interpretation is that it reproduces the phenomenology of the Copenhagen interpretation for macro-systems and the collapse of the wave function is just an innocuous process of removing ergodically inaccessible parts of the epistemic state. However, there is no sharp Heisenberg cut between microscopic systems and macroscopic systems. One can quantify the degree of classicality as the time required to see a transition between two hypothetically macroscopically distinct states; as order $\tau/\Delta$, where $\Delta$ is the magnitude of the inner product of the two states.
A shortened description of new interpretation  will appear in \cite{lett}.

Finally, we can put Schr\"odinger's cat out of it misery. According to the new interpretation, it is either alive or dead but these states are not pure states rather they are associated to ergodic subsets of ontic states that are, for all physically relevant time scales, mutually inaccessable. So the cat may be either alive or dead but in either case in equilibrium with---hence strongly entangled with---the environment.

\begin{acknowledgments} 
I would like to thank Jacob Barandes for a very fruitful exchange of ideas; many of my ideas arose from and are much sharper as a result of our communications. I am supported in part by the STFC grant ST/G000506/1.
\end{acknowledgments}

\end{document}